\documentclass[%
 reprint,amsmath,amssymb,aps,pra,superscriptaddress
]{revtex4-2}

\usepackage{babel}

\usepackage{graphicx}
\usepackage{dcolumn}
\usepackage{bm}
\usepackage{inputenc}
\usepackage{lipsum}
\usepackage{xcolor}
\usepackage{tikz}
\usepackage{quantikz}
\usepackage{enumerate}
\usepackage{float}
\usepackage{wrapfig}
\usepackage{caption,subcaption}
\usepackage{enumitem}   
\usetikzlibrary{arrows.meta,calc} 
\makeatletter
\def\l@en{\l@english}
\makeatother
\usepackage{physics}

\usetikzlibrary{positioning, shapes.geometric, decorations.pathmorphing, calc}

\tikzset{%
    Det/.pic={
        \draw[fill=black] (0.25,0) arc (0:180:0.25) arc  (180:0:0.25 and sin{15}*0.25);
        \draw (0,0) circle [x radius = 0.25, y radius= 0.25*sin{15}];
        \filldraw[color=red!50!black] (0,0) circle [x radius = 0.25/4, y radius = 0.25*sin{15}/4] ;
    },
    Pulse/.pic={
        \fill[thick,domain=-0.5:0.5,smooth, scale = 0.4, draw = black, fill = gray] plot  (\x,{exp(- \x*\x/0.05)});
    },
    client/.style={
        regular polygon,
        regular polygon sides=6,
        minimum size = .5 cm,
        thick,
        draw = black
    },
    server/.style = {
        regular polygon,
        regular polygon sides=4,
        minimum size = 0.75 cm,
        thick,
        draw = black,
        inner sep =0
    },
    entanglement/.style = {
        decorate,
        decoration={coil, amplitude= 4pt, segment length=5 pt}
    }
}

\newlength{\B}
\newlength{\D}
\newlength{\s}
\begin{document}

\preprint{APS/123-QED}

\title{Single-click protocols for remote state preparation using weak coherent pulses}
\author{Janice van Dam}\altaffiliation{These authors contributed equally and are corresponding authors: j.vandam-3@tudelft.nl; emil.hellebek@nbi.ku.dk}
\affiliation{QuTech, Delft University of Technology, Lorentzweg 1, 2628 CJ, Delft, The Netherlands\\
Kavli Institute of Nanoscience, Delft University of Technology, Lorentzweg 1, 2628 CJ, Delft, The Netherlands\\
Quantum Computer Science, EEMCS, Delft University of Technology, Lorentzweg 1, 2628 CJ, Delft, The Netherlands}
\author{Emil R. Hellebek}\altaffiliation{These authors contributed equally and are corresponding authors: j.vandam-3@tudelft.nl; emil.hellebek@nbi.ku.dk}
\affiliation{Center for Hybrid Quantum Networks (Hy-Q), Niels Bohr Institute, University of Copenhagen, Jagtvej 155A, Copenhagen DK-2200, Denmark}
\author{Tzula B. Propp}
\affiliation{QuTech, Delft University of Technology, Lorentzweg 1, 2628 CJ, Delft, The Netherlands\\
Kavli Institute of Nanoscience, Delft University of Technology, Lorentzweg 1, 2628 CJ, Delft, The Netherlands\\
Quantum Computer Science, EEMCS, Delft University of Technology, Lorentzweg 1, 2628 CJ, Delft, The Netherlands}
\author{Junior R. Gonzales-Ureta}
\affiliation{Q*Bird BV, Delftechpark 1, 2628 XJ, Delft, the Netherlands}
\author{Anders S. Sørensen}
\affiliation{Center for Hybrid Quantum Networks (Hy-Q), Niels Bohr Institute, University of Copenhagen, Jagtvej 155A, Copenhagen DK-2200, Denmark}
\author{Stephanie D.C. Wehner}
\affiliation{QuTech, Delft University of Technology, Lorentzweg 1, 2628 CJ, Delft, The Netherlands\\
Kavli Institute of Nanoscience, Delft University of Technology, Lorentzweg 1, 2628 CJ, Delft, The Netherlands\\
Quantum Computer Science, EEMCS, Delft University of Technology, Lorentzweg 1, 2628 CJ, Delft, The Netherlands}

\begin{abstract}
    Remote state preparation (RSP) allows one party to remotely prepare a known quantum state on another party's qubit using entanglement. This can be used in quantum networks to perform applications such as blind quantum computing or long-distance quantum key distribution (QKD) with quantum repeaters. Devices to perform RSP, referred to as a client, ideally have low hardware requirements, such as only sending photonic qubits. A weak coherent pulse source offers a practical alternative to true single-photon sources and is already widely used in QKD. Here, we introduce two new protocols to the previously known protocol for RSP with a weak-coherent-pulse-based device. The known technique uses a double-click (DC) protocol, where a photon from both the server and the client needs to reach an intermediate Bell state measurement. Here, we add to that a single-click (SC) RSP protocol, which requires only one photon to reach the Bell state measurement, allowing for better performance in certain regimes. In addition, we introduce a double-single-click (DSC) protocol, where the SC protocol is repeated twice, and a CNOT gate is applied between the resulting qubits. DSC mitigates the need for phase stabilization in certain regimes, lowering technical complexity while still improving performance compared to DC in some regimes.
    We compare these protocols in terms of fidelity and rate, finding that SC consistently achieves higher rates than DC and, interestingly, does not suffer from an inherently lower fidelity than the DC, as is the case for entanglement generation. Although SC provides stronger performance, DSC can still show performance improvements over DC, and it may have reduced technical complexity compared to SC. Lastly, we show how these protocols can be used in long-distance QKD using quantum repeaters.
\end{abstract}

\maketitle

\section{Introduction}
Remote state preparation (RSP) allows one party to remotely and securely prepare a specific quantum state on another party's qubit, using classical information combined with shared entanglement resources \cite{bennett2001remote}. One party, we call a client, can be a quantum device with low quantum capabilities, specifically, it does not need a quantum memory. In quantum networks, it has potential applications in areas such as blind quantum computing (BQC) \cite{broadbent2009universal}, memory-assisted quantum key distribution (QKD) \cite{schmidt2020memory}, and QKD over repeater nodes \cite{langenfeld2021quantum}.\\
A key motivation in applications such as QKD and BQC is to minimize quantum hardware requirements for clients, allowing them to perform tasks such as measuring \cite{morimae2013blind, fitzsimons2017private, vanDam2024, ekert1991quantum, bennett1992quantum} or sending quantum states \cite{broadbent2009universal, beveratos2002single, garnier2024composably} while offloading complex operations to the server. Other types of clients have been proposed for BQC, like a client that only performs single qubit gates \cite{li2021blind} or even a completely classical client, in combination with multiple non-communicating servers \cite{morimae2013secure, sheng2015deterministic, li2014triple, quan2023verifiable}, though these clients do not allow for QKD.\\
Here, we will study the scenario where clients prepare and send quantum states. While generating single photons remains technologically challenging, weak coherent pulse (WCP) sources offer a practical alternative: they are not perfect single-photon sources as they emit more than one photon with non-zero probability, but they have much lower cost, are simpler to implement and can be easily modulated to the optimal parameters. Despite their limitations, WCP sources are widely used in QKD \cite{hwang2003quantum, lo2005decoy, wang2005decoy,lucamarini2018overcoming, liao2018satellite} due to well-developed techniques that manage multi-photon emissions \cite{scarani2009security}. Recently, WCP-based approaches have also been explored for BQC \cite{dunjko2012blind, garnier2024composably}. \\
In quantum networks, two protocols are widely used for entanglement generation: the double-click (DC) \cite{barrett2005efficient, simon2003robust} and single-click (SC) \cite{campbell2008measurement, humphreys2018deterministic} protocols. DC requires two photons—one from each participating node—to arrive at a Bell state measurement (BSM) station, a scenario often hindered by high losses. SC, on the other hand, requires only one photon to reach the BSM, thereby significantly increasing the success rate, though at the cost of a lower fidelity. Whereas RSP requires entanglement, the entanglement does not need to be stored in both nodes, such that one of the nodes can have lower hardware requirements. When it comes to RSP with a WCP source on the client side, prior work has focused only on the DC protocol \cite{jiang2019remote}. In this study, we explore whether SC can also enhance RSP protocols with a WCP client. To this end, we 
\begin{itemize}
    \item Introduce two novel approaches to performing RSP using a client with a WCP source: single-click (SC), and double-single-click (DSC), for which patents are pending \cite{patentsc, patentdsc}. DSC repeats the SC protocol and uses a controlled-NOT (CNOT) gate aiming to mitigate the need for phase stabilization;
    \item Confirm that SC can indeed achieve higher rates than DC for the same fidelity, consistent with findings from entanglement generation. We find that DSC can also achieve higher rates than DC;
    \item Show that, interestingly, SC and DSC do not inherently suffer from lower fidelity compared to DC, as SC does in entanglement generation, such that SC and DSC have a strict advantage over DC in certain regimes;
    \item Discuss the potential applications of these protocols for BQC and for QKD in repeater networks, where the increased speed  provided by SC and DSC protocols could facilitate longer-distance and more robust quantum communication setups.
\end{itemize}

\section{RSP protocols}\label{sec:analytics}
\begin{figure}
    \begin{subfigure}{0.28\linewidth}
        \begin{tikzpicture}
            \node (Client) [client] {};
            \node (Server) at (0, 2*\B) [server] {\huge .};
            \node (BSM) at (-\B,\B) {};
            \node (Det1) at (-\B-\D,\B+\D) {};
            \node (Det2) at (-\B-\D,\B-\D) {};
            \draw[rotate = 45,transform shape,scale=0.75] pic at (Det1)  {Det};
            \draw[rotate = 135,transform shape,scale=0.75] pic at (Det2)  {Det};
            \draw[rotate=45,transform shape, fill opacity=0.3] pic at ($(BSM)!.7!(Server.south west)$) {Pulse};
            \draw[rotate=45,transform shape,  fill opacity=0.3] pic at ($(BSM)!.3!(Server.south west)$) {Pulse};
            \draw[rotate=135,transform shape, fill opacity=0] pic at ($(BSM)!.7!(Client.north west)$) {Pulse};
            \draw[rotate=135,transform shape, fill opacity=0] pic at ($(BSM)!.3!(Client.north west)$) {Pulse};
            \draw (Client) -- (Det1.center);
            \draw (Server) -- (Det2.center);
        \end{tikzpicture}
        \caption{Double-click}
    \end{subfigure}
    \begin{subfigure}{0.28\linewidth}
        \centering
        \begin{tikzpicture}
            \node (Client) [client] {};
            \node (Server) at (0, 2*\B) [server] {\huge .};
            \node (BSM) at (-\B,\B) {};
            \node (Det1) at (-\B-\D,\B+\D) {};
            \node (Det2) at (-\B-\D,\B-\D) {};
            \draw[rotate = 45,transform shape,scale=0.75] pic at (Det1)  {Det};
            \draw[rotate = 135,transform shape,scale=0.75] pic at (Det2)  {Det};

            \draw[rotate=45,transform shape, fill opacity=0.8] pic at ($(BSM)!.5!(Server.south west)$) {Pulse};
            \draw[rotate=135,transform shape, fill opacity=0] pic at ($(BSM)!.5!(Client.north west)$) {Pulse};
            \draw (Client) -- (Det1.center);
            \draw (Server) -- (Det2.center);
        \end{tikzpicture}
        \caption{Single-click}
    \end{subfigure}
    \begin{subfigure}{0.4\linewidth}
        \begin{tikzpicture}
            \node (Client) [client] {};
            \node (Server) at (0, 2*\B) [server] {\huge ..};
            \node (BSM) at (-\B,\B) {};
            \node (Det1) at (-\B-\D,\B+\D) {};
            \node (Det2) at (-\B-\D,\B-\D) {};
            \draw[rotate = 45,transform shape,scale=0.75] pic at (Det1)  {Det};
            \draw[rotate = 135,transform shape,scale=0.75] pic at (Det2)  {Det};
            \draw (Server.south west) -- (Det2.center);
            \draw (Server.south east) -- (Det2.center);
            \draw[rotate=45,transform shape, fill opacity=0.8] pic at ($(Server.south west)!.5!(Det2)$) {Pulse};
            \draw[rotate=35,transform shape, fill opacity=0.8] pic at ($(Server.south east)!.3!(Det2)$) {Pulse};
            \draw[rotate=135,transform shape, fill opacity=0] pic at ($(BSM)!.7!(Client.north west)$) {Pulse};
            \draw[rotate=135,transform shape, fill opacity=0] pic at ($(BSM)!.3!(Client.north west)$) {Pulse};
            \draw (Client) -- (Det1.center);
        \end{tikzpicture}
        \caption{Double-single-click }
    \end{subfigure} 
    \caption{Diagram illustrating the three remote state preparation protocols. The square represents the server, with a dot denoting an emitting qubit, while the hexagon represents the client equipped with a weak coherent pulse source. Both client and server send their quantum states to the Bell state measurement station. In the double-click protocol in (a), a two-mode encoding is used, such as polarization or time-bin, while in the single-click (b) and double-single-click (c) protocols, presence-absence encoding is applied in a single-mode configuration. The protocols in (a) and (c) both require two photons to be detected, but the protocol (c) allows for re-trying of each node separately, as the photons are entangled with separate memory qubits. Two lines are drawn in (c) to illustrate the pulses coming from the two different memory qubits, it is not required to have two separate paths to the BSM. The server’s qubit is entangled with the emitted light, which, along with the client’s light, is mixed on a beam-splitter and analyzed by the BSM. Upon successful detection, the server's qubit is projected into a state selected by the client.}
    \label{fig:Schematics}
\end{figure}
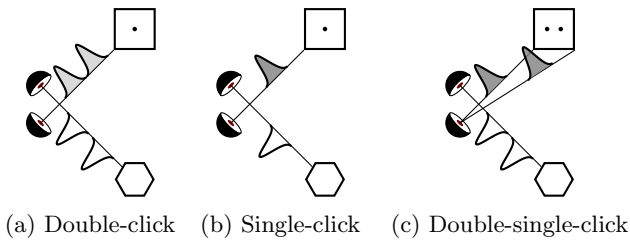

\begin{figure*}
    \centering
    {\centering
     \subfloat[]{\includegraphics[width=0.49\textwidth]{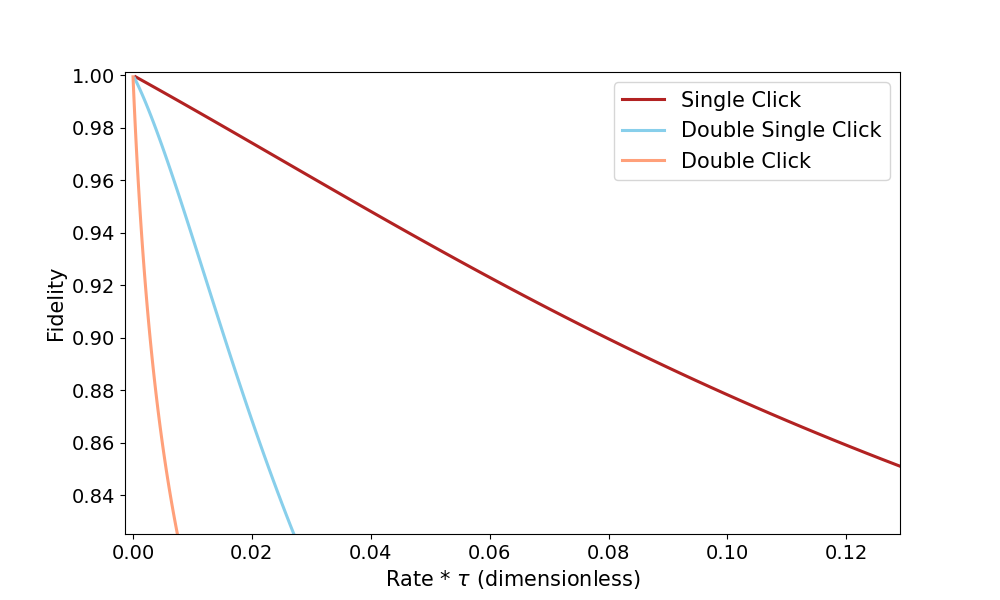}}}
    {\centering
     \subfloat[]{\includegraphics[width=0.49\textwidth]{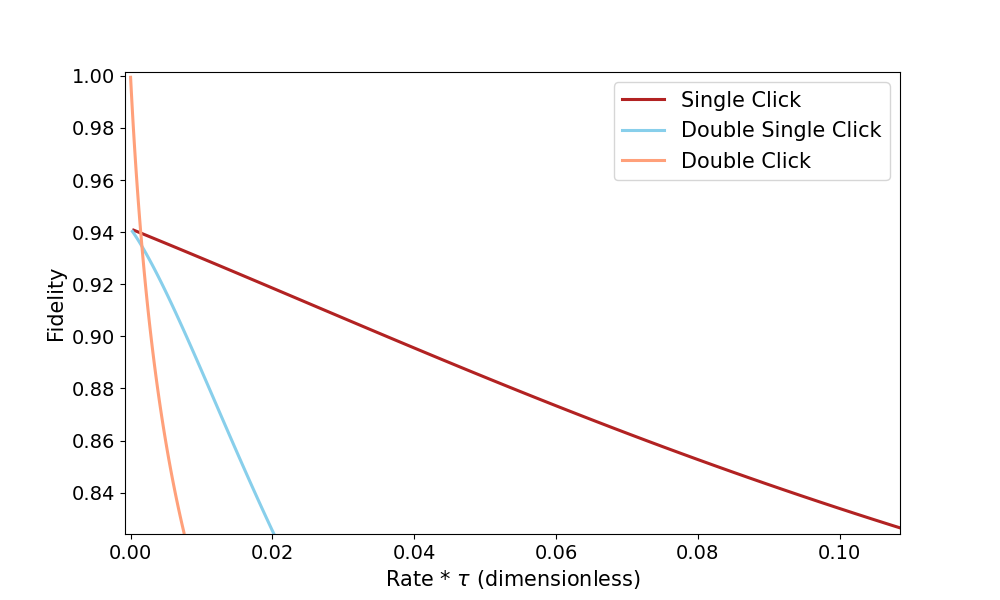}}}
    \caption{Trade-offs between fidelity and rate for the three RSP protocols, (a) without phase noise and (b) with phase noise of $\sigma_\text{SC}=\sigma_\text{DSC}=0.5$ rad. This analysis assumes a server efficiency of $\eta_s=0.13$ and a detector efficiency of $\eta_d=0.7$. The plotted values are for a mean photon number, as emitted by the client, below 0.5.}\label{fig:RF}
\end{figure*}

We consider three protocols for performing RSP between a client equipped with a WCP source, and a quantum server, equipped with quantum memory that can emit memory-entangled photons. In all cases a BSM station is required between the client and the server. Mathematically speaking, the exact position of the BSM (in the middle, closer to client or closer to the server) is irrelevant, as the fiber losses can be taken into account by adjusting the efficiencies $\eta_c$, $\eta_s$. However, positioning the BSM directly next to the server has some advantages: it reduces the number of nodes needed in the network and it allows for faster RSP attempts. As in most regimes we are limited by the repetition rate of the server, having the BSM close to the server allows for a reduction of classical communication time. The server can quickly receive success/failure signals, and reset for the next attempt, while the client can continuously send pulses at a high rate.\\
We refer to the three protocols as single-click (SC), double-single-click (DSC) and double-click (DC), a schematic overview of the protocols is provided in Figure \ref{fig:Schematics}. This terminology is analogous to that used in entanglement generation, where the DC protocol uses orthogonal modes such as polarization or time-bin encoding to carry the quantum information of the photons. The SC protocol uses a presence-absence encoding, allowing entanglement to be formed with only one photon arriving at the BSM \cite{campbell2008measurement}. The DSC protocol repeats the SC protocol to eliminate the need for phase stabilization in certain regimes, a topic discussed further in Section \ref{subsec:phase_errors}. \\
Here, we carry over these known techniques in entanglement generation to RSP using a WCP source. While the DC protocol in RSP is not novel \cite{jiang2019remote}, we provide its description and analytical expressions for the rate and fidelity we can achieve using this protocol here and in Appendix \ref{app:derivation} for completeness and comparison.\\
Sections \ref{subsec:scanalytics}, \ref{subsec:dscanalytics}, \ref{subsec:dcanalytics} cover each RSP protocol, along with analytic performance results, which are plotted in Figure \ref{fig:RF} and discussed in Section \ref{sec:comp+disc}. A detailed derivation of the analytics can be found in Appendix \ref{app:derivation}. Afterwards, in Section \ref{subsec:phase_errors} we will provide an extension to the SC and DSC analytics to include errors due to phase noise.\\
Apart from phase noise and losses, we assume ideal hardware, thus excluding decoherence, gate imperfections, and infidelity in the server photon-matter state. The aim here is to capture the errors that are inherent to using WCPs instead of a single photon source and the differences between the protocols.
\subsection{Model}
Each calculation below will follow the same steps. First, we characterize the input states from the client and server.
Here, we look at creating states on the equator of the Bloch sphere, i.e., states of the form $\ket{+_\theta}=(\ket{0} + e^{i\theta}\ket{1})/\sqrt{2}$ for some angle $\theta$, which are states sufficient for e.g. QKD \cite{ma2012phasemdi} an BQC \cite{kapourniotis2023asymmetric}.\\
A loss channel with loss probability $1-\eta_c$ and 1-$\eta_s$ is applied to the client and server state, respectively. These losses capture all losses and inefficiencies in the system: emission inefficiency, coupling to fiber, inefficiencies due to frequency conversion, loss in fiber and detector inefficiencies. After, we consider the photon(s) arriving at the beamsplitter of the BSM, where beam splitter transformations transformations are applied to the combined client-server state. Then, projection operators are applied for the measurement, allowing us to find the density matrix of the final remotely prepared state as well as the probability of this measurement outcome. From this, we construct the (dimensionless) rate per attempt time $\tau$ (in SC and DC equal to the success probability of an attempt) and the fidelity with respect to the target state $\ket{+_\theta}$, as $F=|\bra{+_\theta}\rho\ket{+_\theta}|$.

\subsection{Double-click protocol}\label{subsec:dcanalytics}
In the DC protocol, the server prepares a Bell state, consisting of the server qubit and the server photon, the latter encoded in e.g. time-bin or polarization. The client emits a WCP with amplitude $\alpha=\abs{\alpha}$ in equal superposition of the photon encoding, with a relative phase $\theta$. The phase of the WCP pulse is assumed to be randomized over time for security \cite{garnier2024composably}. If the BSM detects a photon in both states of the encoding, the operation is successful and $\ket{+_\theta}=(\ket{0} + e^{i\theta}\ket{1})/\sqrt{2}$ is prepared on the server. The relative phase between the WCPs determines the phase of the prepared state. With this, we compute the fidelity of the remotely prepared state, and the RSP success probability as the rate times the time per attempt $\tau$ and find
\begin{equation}\label{eq:F_DC}
\begin{split}
    &F_\text{DC} = \frac{1}{2}\left[1+\frac{\eta_s\eta_c\abs{\alpha}^2/8}{\pqty{1-e^{-\eta_c\abs{\alpha}^2/4}}} \right. \times\\
    &\left. \frac{1}{\bqty{\frac{\eta_s}{2}\pqty{1+\frac{\eta_c\abs{\alpha}^2}{4}}+\pqty{1-e^{-\eta_c\abs{\alpha}^2/4}}\pqty{1-\eta_s}}}\right]\\
    &\sim 1-\frac{\eta_c}{\eta_s}\frac{4-3\eta_s}{16}\abs{\alpha}^2,
\end{split}
\end{equation}
\begin{equation}\label{eq:P_DC}
\begin{split}
    &P_\text{DC} =R_\text{DC}\tau=\\
    &4e^{-\eta_c\abs{\alpha}^2/2}\pqty{1-e^{-\eta_c\abs{\alpha}^2/4}} \times\\
    &\bqty{\frac{\eta_s}{2}\pqty{1+\frac{\eta_c\abs{\alpha}^2}{4}}+\pqty{1-e^{-\eta_c\abs{\alpha}^2/4}}\pqty{1-\eta_s}}\\
    &\sim \frac{\eta_c\eta_s}{2} \abs{\alpha}^2.
\end{split}
\end{equation}
In the last lines we take the approximation $|\alpha|^2\ll 1$, and leave the derivation of the protocol to be found in Appendix \ref{app:derivation}.\\
From this we can see that the fidelity approaches one as the mean photon number approaches zero, which we also see in Figure \ref{fig:RF}. The mean photon number approaching zero, however, also means that the probability of success approaches zero, as no photons get send to the BSM. As we increase the mean photon number, the probability of a successful event increases, but with that, the fidelity drops due to the increased probability of multi-photon events. This is true for all three protocols in the absence of phase noise.

\subsection{Single-click protocol}\label{subsec:scanalytics}
In the SC protocol, the client sends out a coherent state with complex amplitude $\alpha=\abs{\alpha}e^{-i\theta}$. The server is initially in a superposition state $\sqrt{1-\xi^2}\ket{0}+\xi\ket{1}$, where we refer to $\xi$ as the bright state parameter. The server sends out a single photon if its memory qubit is in the bright state, denoted $\ket{1}$, and no photon otherwise, denoted $\ket{\emptyset}$. With that, the input states are described by
\begin{align}\label{eq:input_c}
    \ket{\psi_c} &= \ket{\alpha}=e^{-\abs{\alpha}^2/2} \sum_n \frac{\alpha^n}{\sqrt{n!}} \ket{n},\\\label{eq:input_s}
    \ket{\psi_s} &= \sqrt{1-\xi^2} \ket{\emptyset 0}+\xi \ket{11},
\end{align}
where $\ket{\psi_c}$ ($\ket{\psi_s}$) is the state emitted by the client (server).\\
The state remotely prepared at the server depends on two factors: (1) The balance of probabilities for a photon arriving from each side, where equal probabilities create a state on the equator of the Bloch sphere \footnote{This is because, the photon coming from the client leaves the server qubit in $\ket{0}$, because the server did not emit a photon and thus was not in the bright state, and the photon coming from the server leaves the qubit in $\ket{1}$, equal probabilities give a state with equal weights $\ket{0}$ and $\ket{1}$, hence on the equator of the Bloch sphere.}. These probabilities are determined by the losses on the server side $\eta_s$, the losses on the client side $\eta_c$, the mean photon number of the WCP, $|\alpha|^2$, and the bright state parameter $\xi$ of the emitter at the server. (2) The phase $\theta$ of the WCP, since the state coming from the server has a constant phase (assumed to be zero for simplicity), this introduces a phase difference of $\theta$ between the states produced when a photon originating from the server side (heralding the server in the bright state) and the client side (heralding the server in the dark state) arrives at the BSM.
\\
 To create our target state $\ket{+_\theta}$, we need to have equal probability of the photon arriving at the BSM from the client, as from the server. Note that, unlike for DC, we can only prepare states on a fixed latitude of the Bloch sphere (here, the equator); changing latitude requires either the server to change its bright state parameter, or the client to change the laser intensity. While this could in principle be done, we omit this possibility for simplicity.\\
After applying the loss channels, beam splitter transformations and detector projectors to the combined client-server state, we compute the fidelity of the remotely prepared state with respect to the target state $\ket{+_\theta}$, optimized over $\xi$, and the RSP success probability as the rate times the time per attempt $\tau$ and find \\
\begin{equation}\label{eq:F_SC}
\begin{split}
    &F_\text{SC}=\frac{1}{2}\left[1+\sqrt{\frac{\eta_c\eta_s\abs{\alpha}^2/2}{\pqty{1-e^{-\eta_c\abs{\alpha}^2/2}}}}\right. \times
    \\&\left.\sqrt{\frac{1}{2\pqty{1-\eta_s}\pqty{1-e^{-\eta_c\abs{\alpha}^2/2}}+\eta_s\pqty{1+\eta_c\abs{\alpha}^2/2}}}\right]\\&\sim 1-\frac{\eta_c(4-3\eta_s)}{16\eta_s}\abs{\alpha}^2,
\end{split}
\end{equation}
\begin{equation}\label{eq:P_SC}
\begin{split}
    &P_\text{SC} = R_\text{SC}\tau=\\
    &2e^{-\frac{\eta_c\abs{\alpha}^2}{2}}\pqty{1-e^{-\frac{\eta_c\abs{\alpha}^2}{2}}}\times\\
    &\bqty{1+\frac{\eta_s\pqty{2e^{-\frac{\eta_c\abs{\alpha}^2}{2}}-1+\frac{\eta_c\abs{\alpha}^2}{2}}}{\eta_s\pqty{2e^{-\frac{\eta_c\abs{\alpha}^2}{2}}-1+\eta_c\abs{\alpha}^2}+4\pqty{1-e^{-\frac{\eta_c\abs{\alpha}^2}{2}}}}}\\
    &\sim 2\eta_c\abs{\alpha}^2.
\end{split}
\end{equation}
In the approximation on the last line we assume $|\alpha|^2\ll 1$. A detailed derivation can be found in Appendix \ref{app:derivation}.\\
In Figure \ref{fig:RF} (a), in red we see the rate-fidelity trade-off for the SC protocol. Just like for DC, we see that the fidelity approaches one as the rate approaches zero, which happens for the mean photon number $|\alpha|^2\rightarrow 0$. The fidelity in equations \ref{eq:F_DC} and \ref{eq:F_SC} are both a linear function of $|\alpha|^2$, but the fidelity of DC has twice the negative slope of SC in the small $|\alpha|^2$ limit. Combined with a different scaling in the rate, we find a more favorable rate-fidelity trade-off for SC, characterized by a less-steep slope in Figure \ref{fig:RF} (a). 

\subsection{Double-single-click protocol}\label{subsec:dscanalytics}
\begingroup
\interlinepenalty=0     
\predisplaypenalty=0    
The DSC protocol involves performing the SC protocol twice, thus remotely preparing two qubits as described in the previous section. We assume here that the first qubit is not affected during the generation of the second qubit, meaning that decoherence is not included, giving an upper bound to the achievable fidelity. After successfully heralding two clicks, the resulting qubits are in a state where either zero ($i\ket{00}$), one ($\ket{01}$ or $\ket{10}$) or two ($\ket{11}$) photons have been emitted by the server emitter. A CNOT gate is then applied to the two remotely prepared qubits, followed by a measurement of the target qubit. We post-select on the target state being in $\ket{0}$, excluding state where an even number of photons have been emitted, similar to the DC protocol. Then, the phase of the final qubit corresponds to the difference of the first two qubits, as depicted in Figure \ref{fig:CNOTcirc}. With this, we are thus able to mitigate the need for phase stabilization if the phase stays constant between the two emissions, this is discussed in Section \ref{subsec:phase_errors}. A detailed analysis of the protocol leads to the expressions for fidelity and rate
\endgroup
\begin{equation}\label{eq:F_DSC}
\begin{split}
    &F_\text{DSC} = \frac{1} {2}\left[ 1+\frac{\eta_s\eta_c\abs{\alpha}^2/4}{\pqty{1-e^{-\frac{\eta_c\abs{\alpha}^2}{2}}}}\right.\times\\
    &\left. \frac{1}{\bqty{\frac{\eta_s}{2}\pqty{1+\frac{\eta_c\abs{\alpha}^2}{2}}+\pqty{1-\eta_s}\pqty{1-e^{-\frac{\eta_c\abs{\alpha}^2}{2}}}}
    }\right]\\
    &\sim 1-\frac{\eta_c}{\eta_s}\frac{4-3\eta_s}{8}\abs{\alpha}^2,
\end{split}
\end{equation}
\begin{equation}\label{eq:P_DSC}
\begin{split}
    &R_\text{DSC}\tau = \frac{8\left(1-e^{-\frac{\eta_c|\alpha|^2}{2}}\right)}{3\eta _s^2e^{\frac{\eta_c\abs{\alpha}^2}{2}}} \times\\
    &\frac{\bqty{\eta _s \pqty{3+\eta_c|\alpha|^2}+4\pqty{1-e^{-\frac{|\alpha|^2 \eta _c}{2}}} \pqty{1-\eta _s}}}{\bqty{3+\eta_c|\alpha|^2-4\pqty{1-e^{-\frac{\eta_c|\alpha |^2}{2}}}}^2}\times\\
    &\bigg\{8-\eta _s \left(1-2\eta_c|\alpha|^2\right) -4e^{-\frac{\eta_c|\alpha |^2}{2}} \left(2-\eta _s\right)\\
    &\quad-4
   \sqrt{\left(1-e^{\frac{-\eta_c| \alpha | ^2}{4}}\right)}\times\\
   &\sqrt{\eta _s \left(3+\eta_c|\alpha|^2\right)+4\pqty{1-e^{\frac{-\eta_c| \alpha | ^2}{2}}}
   \left(1-\eta _s\right)}\bigg\}\\
   &\sim \frac{4}{3}\eta_c\abs{\alpha}^2.
\end{split}
\end{equation}

We again approximate for $|\alpha|^2\ll 1$, and the derivation is found in Appendix \ref{app:derivation}. \\
In Figure \ref{fig:RF} (a) we see that the blue line for DSC lies between those for DC and SC, as we see the same scaling of $F$ with $|\alpha|^2$ as for DC (Equations \ref{eq:F_DC} and \ref{eq:F_DSC}), but a different scaling in the rate (Equations \ref{eq:P_DC} and \ref{eq:P_DSC}).

\subsection{Phase noise}\label{subsec:phase_errors}
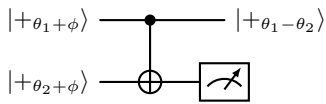
\begin{figure}\label{fig:CNOTcirc}
    \centering
    \begin{quantikz}
        \lstick{$\ket{+_{\theta_1+\phi}}$} & \ctrl{1} &  \rstick{$\ket{+_{\theta_1-\theta_2}}$} \\
        \lstick{$\ket{+_{\theta_2+\phi}}$}& \targ{} & \meter{}  
    \end{quantikz}
    \caption{Local operations in the DSC protocol: some random phase $\phi$ that gets added through phase wandering gets canceled out after applying a CNOT gate and measuring the target qubit, as long as the random phase is the same for the two qubits.}
    \label{fig:CNOTcirc}
\end{figure}

Small fluctuations in the system cause the true phase of the WCP to drift over time. In the DC protocol, with time-bin or polarization encoding, these phase shifts are not problematic because it is only sensitive to phase differences between two co-propagating orthogonal modes. Typically, these modes are close enough in time or have no fluctuating birefringence, so that any phase changes between the two modes are negligible. In contrast, phase shifts present a challenge for the SC protocol, necessitating active phase stabilization, as demonstrated in, e.g., \cite{stolk2024extendable}. The DSC protocol aims to eliminate this technical overhead by ``deleting" the overall phase of two successful SC operations, leaving only the phase \textit{difference} between the two clicks as noise. We note that with that, DSC does not combat phase noise, instead it combats the technical requirement of phase stabilization. The CNOT and it's effect on the qubits is depicted in Figure \ref{fig:CNOTcirc}. This approach is effective only if the phase drift between the two successes is small, which requires sufficiently fast remote preparation.\\
Despite phase stabilization in SC and the CNOT operation in DSC, some residual phase noise will likely appear in both protocols. In SC, this may arise from imperfections in the phase stabilization process. In DSC, the phase noise stems from drift between the first and second clicks, which depends on the time interval between them ($T$) and the linewidth of the optical field and drift of optical elements ($\Delta \nu$), i.e., the rate at which the phase evolves. We model phase noise in SC and DSC as affecting the final state in similar ways, but to differing extents. To account for this, we introduce two variables to represent the standard deviation of phase noise: $\sigma_\text{SC}$ for SC and $\sigma_\text{DSC}$ for DSC.\\
The value of $\sigma_\text{SC}$ can be determined through experimental characterization of the setup, while $\sigma_\text{DSC}$ can be estimated using $\sqrt{2\pi \Delta \nu T}$ \cite{czegledi2016polarization}, if limited by optical linewidth.\\
We assume phase noise follows a Gaussian distribution with standard deviation $\sigma$. This results in a phase noise factor of $X_\text{noise}=e^{-\sigma^2/2}$, which modifies the fidelity equations for SC and DSC. Without noise, these fidelities, given in Equations \ref{eq:F_SC} and \ref{eq:F_DSC}, take the form $F=(1+x)/2$; with noise, they become $F_\text{noise}=(1+xX_\text{noise})/2$. Additional details are provided in Appendix \ref{subapp:phasenoise}.\\
In Figure \ref{fig:RF} (b) we see the effect of the phase noise term on the SC and DSC results. While the rate of the protocols is not affected by the phase noise, it lowers the maximal achievable fidelity, effectively pushing the rate-fidelity line downwards. Here, we have chosen to plot $\sigma_\text{SC}=\sigma_\text{DSC}$ for comparison. These two parameters will likely not be the same in a real-life system, as the noise has a different origin for each protocol.
\begin{figure*}[!ht]
\centering
    {\centering
     \subfloat[Protocol that gives the highest rate given a target fidelity of 0.98 for $\sigma_\text{DSC}=0.5$ rad.]{\includegraphics[width=0.33\textwidth]{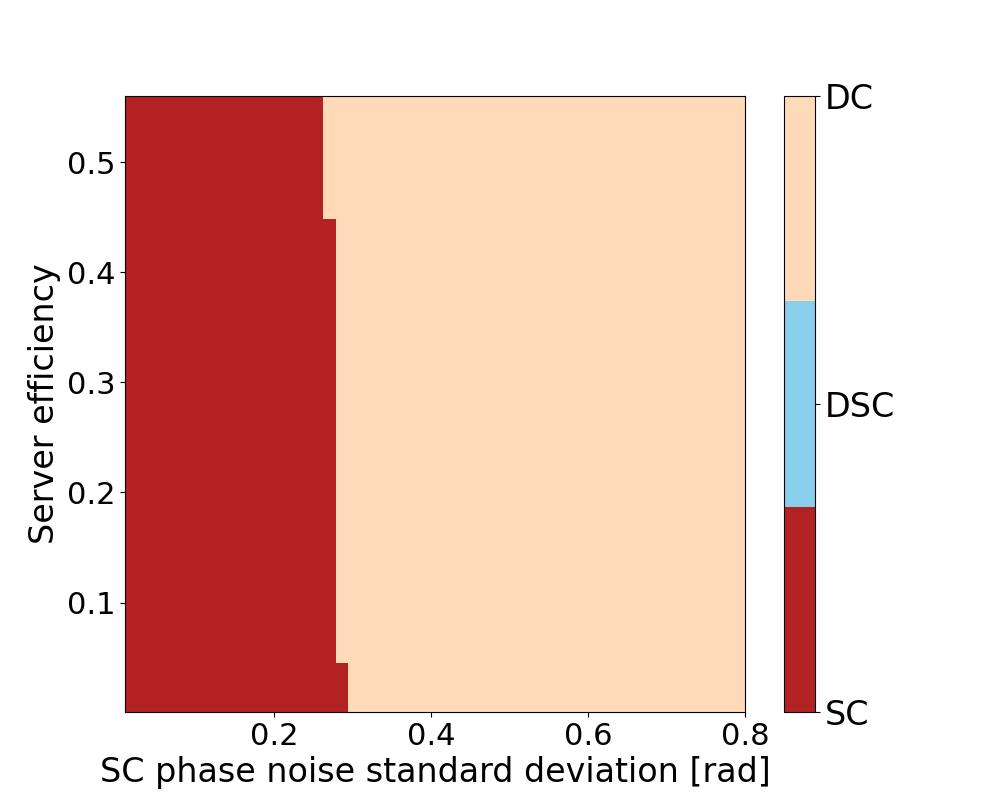}}}
    {\centering
     \subfloat[Protocol that gives the highest fidelity given a target rate of 0.01$\tau^{-1}$ for $\sigma_\text{DSC}=0.5$ rad.]{\includegraphics[width=0.33\textwidth]{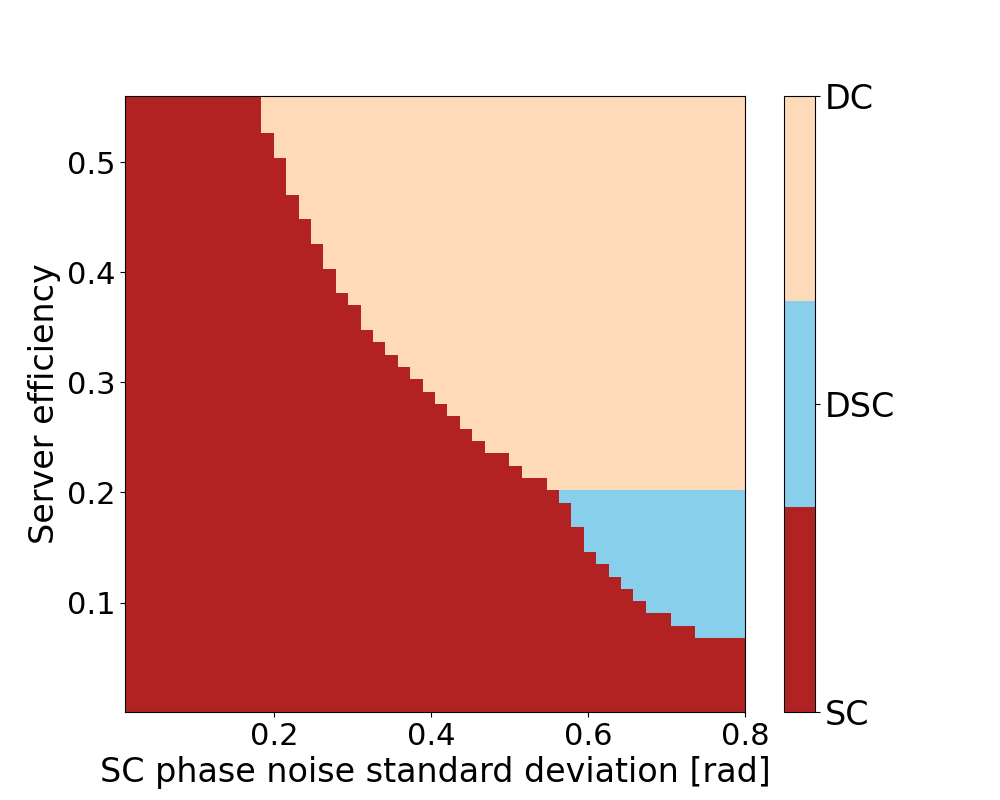}}}
    {\centering
     \subfloat[Protocol that gives the highest fidelity given a target rate of 0.01$\tau^{-1}$ for $\eta_s=0.3$.]{\includegraphics[width=0.32\textwidth]{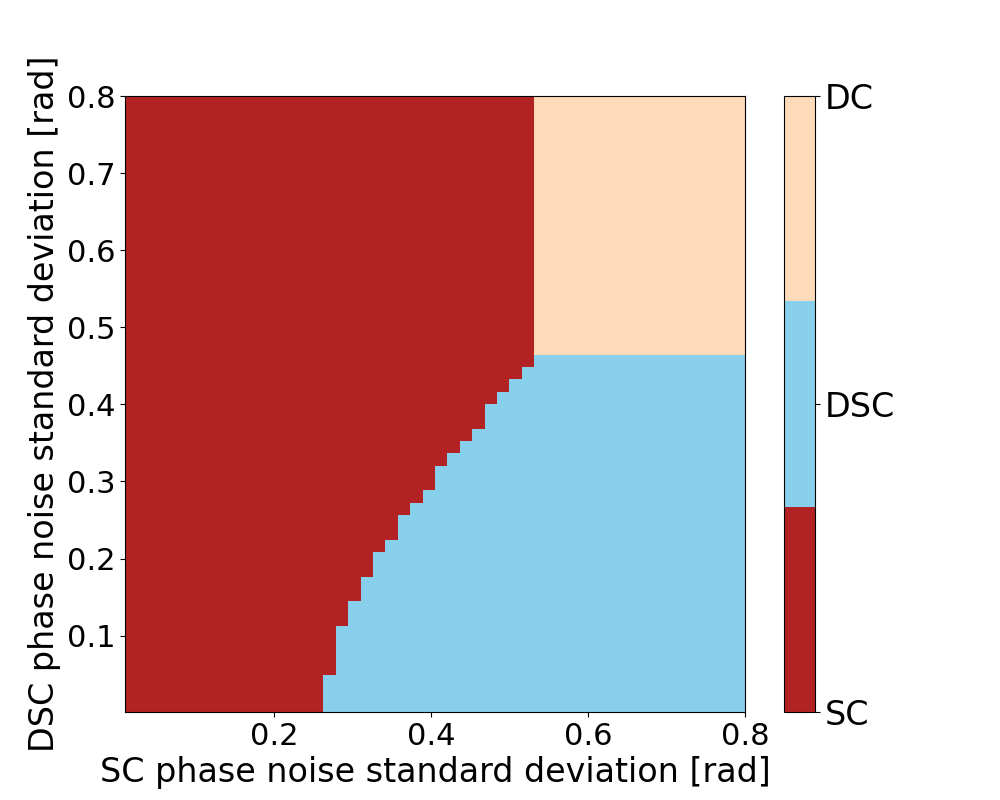}}}
    \caption{Optimal RSP protocol — DC, DSC, or SC — for a given target fidelity or rate, across varying levels of phase noise $\sigma_\text{SC}$ and $\sigma_\text{DSC}$ (standard deviation in radians), and server efficiency. For each combination of parameters, we determine the smallest $|\alpha|^2$ at which the target is met (if achievable) and identify the protocol that performs best. Panels show: (a) the protocol providing the highest rate for a target fidelity of 0.98, while $\sigma_\text{DSC}$ is set to 0.5 rad, (b) the protocol providing the highest fidelity for a target rate of 0.01, again with $\sigma_\text{DSC}=0.5$ rad, and (c) the protocol providing the highest fidelity for a target rate of 0.01, for a fixed server efficiency of $\eta_s = 0.3$ while varying $\sigma_\text{DSC}$. The best-performing protocol is represented by color: DC (peach), DSC (blue), and SC (red).}\label{fig:best}
\end{figure*}
\section{Comparison and discussion}\label{sec:comp+disc}
From the formulas for the fidelity without phase noise, Equations \eqref{eq:F_DC}, \eqref{eq:F_SC} and \eqref{eq:F_DSC}, we note that the first order contributions in $|\alpha|^2$ are $1-\eta_c(4-3\eta_s)|\alpha|^/16\eta_s$ for SC and DC and $1-\eta_c(4-3\eta_s)|\alpha|^/8\eta_s$ for DSC. This infidelity can be intuitively understood as a client photon reaching the detector in the same mode as the server photon. Either the server photon is lost, contributing with $1-\eta_s$, or the server photon reaches the detector and bunches with the client photon, contributing with $\eta_s/4$. In the DSC protocol there are two chances to introduce this error, leading to an extra factor of 2. For the small values of $|\alpha|^2$ considered in this paper, the fidelity will thus be very similar across the three protocols. However, the performance of the protocols depends heavily on the rate and the phase noise.\\
Thus, to compare the performance of the three protocols, we look at their fidelity and rate in different scenarios. We do this by varying the intensity of the client laser $\alpha$, the amount of phase noise in the SC and DSC protocols $\sigma_\text{SC}$ and $\sigma_\text{DSC}$ and the efficiency of the server $\eta_s$. Other parameters that we consider are the efficiency on the client side, $\eta_c$, determined mostly by fiber loss, and the single photon detector efficiency $\eta_d$, however we do not vary over these. We consider a scenario where the client and server are separated by 25 km of optical fiber, resulting in a transmission probability of $\eta_c = 0.32$ for the client pulse to reach the BSM, which is located at the server. The same rates and fidelities can be achieved for other distances, by simply adjusting $\alpha$, for this reason, $\eta_c$ is not considered as a parameter to vary, but it is noted as it affects the values of $\alpha$. Adjusting $\alpha$ will affect the security, and might thus lead to an overhead on the full (QKD or BQC) protocol. This effect is not considered here. Similarly, we set $\eta_d=0.7$, as varying the detection efficiency has the same effect as varying $\eta_s$ and $\eta_c$, which does not give use new insights in trade-offs.\\
We see that increasing $\alpha$ enhances the rate but reduces the fidelity, leading to a trade-off explored for the three different protocols in Figure \ref{fig:RF}. For this figure, we set the efficiency of the server to $\eta_s=0.13$ based on an emission probability of 0.53 \cite{schupp2021interface} and a successful telecom frequency conversion probability of 0.25 \cite{krutyanskiy2019light}, representing the state-of-the-art. We look at both a scenario with no phase noise in \ref{fig:RF}(a) and additionally, we look at what happens when we introduce a phase noise of a standard deviation of $\sigma_\text{SC}=\sigma_\text{DSC}=0.5$ radians \cite{stolk2024extendable} for both SC and DSC in Figure \ref{fig:RF} (b). With this, we vary the laser intensity of the client between $|\alpha|^2=0.001$ and $|\alpha|^2=0.5$.\\

From Figure \ref{fig:RF}(a) we see that the SC protocol can achieve the highest rates, followed by DSC, with DC having the lowest rate for any $\alpha$. This was expected and is analogous to advantages known for SC in entanglement generation. However, unlike in entanglement generation or when using on;y single photons, the SC and DSC protocols provide similar fidelity for a given $|\alpha|^2$ to DC which gives it a strict advantage over DC in the absence of phase noise.\\
From Figure \ref{fig:RF}(b) we see that in the regime of low rates, the DC protocol can obtain a higher fidelity. This is due to phase noise lowering the maximally achievable fidelity for SC and DSC. In a scenario without phase noise (Figure \ref{fig:RF}(a)), the fidelity approaches unity for all protocols. To achieve a reasonable rate, the  laser power must, however, be increased. This higher $\abs{\alpha}^2$ causes an increase in the infidelity of the three protocols, with the greatest impact on the DC protocol, as there $\abs{\alpha}^2$ needs to be increased the most for the same increase in rate. For SC, the required increase in $\abs{\alpha}^2$, and thereby the increase in infidelity, is the lowest. Thus, for a specific rate, the SC protocol can yield a higher fidelity than the DC protocol. 
\\
To determine which protocol is most advantageous under various conditions, we evaluate scenarios across different server efficiencies (factoring in detector efficiency: $\eta_s \rightarrow \eta_s \eta_d$) and levels of phase noise for specific target fidelities or rates. Figure \ref{fig:best} illustrates these findings, with blocks color-coded to 
indicate the most advantageous protocol.\\
Figure \ref{fig:best}(a) shows the regions where each scheme provides the highest rate for a target fidelity of 0.98. Here, we have fixed the phase noise of the DSC protocol to be $\sigma_\text{DSC}=0.5$, but we vary $\eta_s$ and $\sigma_\text{SC}$. SC outperforms DC in low-$\sigma_\text{SC}$ regimes, with occasional jumps based on server efficiency. This is because, in high-$\sigma_\text{SC}$ regions, SC and DSC fail to achieve the target fidelity, leaving DC as the only viable option. Near the boundary between the SC and DC regions, SC approaches its maximum fidelity and achieves the target at a lower rate than DC. Here, higher server efficiencies favor DC more significantly than SC, further shrinking SC's advantageous region. However, unless the target fidelity is close to the SC maximum, SC generally provides higher rates than DC.\\
The size of SC's advantage region decreases with increasing target fidelity and increases with lower target fidelity, as the target fidelity determines where SC becomes infeasible. DSC does not exhibit a clear advantage in any region, but appears on the boarder between the SC and DC regions when $\sigma_\text{DSC}\leq 0.3$ rad. However, as we will also discuss later, DSC provides an advantage over SC in terms of technical demand in the form of phase stabilization.\\
In Figure \ref{fig:best}(b), we identify the scheme achieving the highest fidelity for a target rate of 0.01, again setting $\sigma_\text{DSC}=0.5$. DC outperforms SC and DSC only in scenarios with high $\sigma_\text{SC}$ and efficient servers. As the target rate decreases, the size of DC’s advantage region increases, while it decreases for higher target probabilities. DSC gains some advantage over DC and SC when the server efficiency is low, but $\sigma_\text{SC}$ is high. Naturally, this region gets smaller when $\sigma_\text{DSC}$ is higher and larger when $\sigma_\text{DSC}$ is lower.\\
To find the effect that different ratios of $\sigma_\text{SC}$ to $\sigma_\text{DSC}$ has on the advantage regions, we fix $\eta_s$ to 0.3 and again find which protocol provides the highest fidelity given a target rate of 0.01, these regions are shown in Figure \ref{fig:best}(c). Here, we clearly see the divide between SC and DSC being advantageous depending on the noise levels occurring in each protocol. The high-noise regime in which DC provides an advantage shrinks when the server efficiency is lower, and grows when it is higher. Additionally, increasing the server efficiency pushes down, towards lower $\sigma_\text{DSC}$, the border line between SC and DSC, and decreasing it will push the line up.\\
Alternatives to the selected values for these plots are given in Appendix \ref{app:altplots}, which can be used to visualize the explained effects of parameter changes of the form \textit{when parameter x increases/decreases, y happens}.\\
Overall, DC is advantageous under high phase noise, high server efficiency, or high fidelity targets. SC performs better with less efficient servers, lower phase noise, or when high rates are desired. DSC can gain an advantage if $\sigma_\text{DSC}$ is small compared to $\sigma_\text{SC}$. Notably, protocols requiring multiple remotely prepared qubits (e.g., in BQC) may benefit more from higher-rate, lower-fidelity RSPs due to the impact of decoherence.\\
A drawback of the SC protocol is the need for phase stabilization, which can be technologically challenging to implement. The DSC protocol might alleviate this need, when two consecutive successes of the SC protocol are close enough in time so that the phase has not shifted significantly, otherwise $\sigma_\text{DSC}$ will grow too large. Therefore, one needs to be able to either re-excite the memory qubit very fast, or have enough multiplexing capabilities. Notably, even if DSC does not give advantageous performance over SC, eliminating the need for phase stabilization might still make it favorable, as long as $\sigma_\text{DSC}$ is low enough to provide an advantage over DC. DC always has the advantage of not needing phase stabilization and also having a negligible amount of phase noise, though the increased rate of SC and DSC over DC can lead to an overall higher fidelity for multi-qubit states when in the presence of decoherence. \\
Some noise sources have been left out of this analysis. For example, infidelity in the server matter-photon state and decoherence of the server memory. For a single RSP in SC and DC, decoherence will be minimal, because the BSM is assumed to be directly next to the server, and thus the RSP will be heralded almost instantaneously. However, in the DSC protocol, decoherence will affect the fidelity of the final remotely prepared state if the time between the two SC successes is long. This effect is not taken into account here, as it would take many assumptions on the setup (e.g., server repetition rate, multiplexing capabilities, cutoff time for the memory) in order to quantify this, making it a less generally applicable comparison. Therefore, the fidelity given here for the DSC protocol can be taken as an upper bound for the achievable fidelity with decoherence. To a certain extent, the amount of infidelity decoherence adds to the final remotely prepared state can be adjusted while sacrificing on rate by setting a window, sometimes referred to as a cutoff time, for the two clicks \cite{davies2024tools}.

\section{Applications}\label{sec:applications}
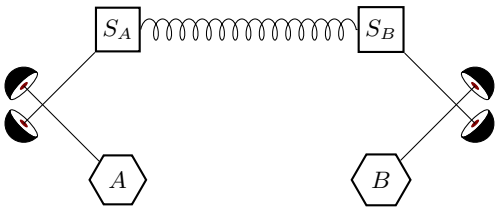
\begin{figure}
    \centering
     \begin{tikzpicture}
        \node (Alice) at (-\B-\s,0) [client] {$A$};
        \node (BSMA) at (-2*\B-\s,\B) {};
        \node (DetA1) at (-2*\B-\s-\D,\B+\D) {};
        \node (DetA2) at (-2*\B-\s-\D,\B-\D) {};
        \node (SAlice) at (-\B-\s, 2*\B) [server] {$S_A$};
    \node (SBob) at (\B+\s, 2*\B) [server] {$S_B$};
    \node (BSMB) at (2*\B +\s, \B) {};
    \node (DetB1) at (2*\B+\s+\D, \B+\D) {};
    \node (DetB2) at (2*\B+\s+\D, \B-\D)  {};
    \node (Bob)  at (\B+\s,0) [client] {$B$};
    \draw[rotate = 45,transform shape] pic at (DetA1)  {Det};
    \draw[rotate = 135,transform shape] pic at (DetA2)  {Det};
    \draw[rotate = -45,transform shape] pic at (DetB1)  {Det};
    \draw[rotate = -135,transform shape] pic at (DetB2)  {Det};

    \draw[entanglement] (SAlice) -- (SBob);
    \draw (Alice) -- (DetA1.center);
    \draw (SAlice) -- (DetA2.center);
    \draw (Bob) -- (DetB1.center);
    \draw (SBob) -- (DetB2.center);
\end{tikzpicture}
    \caption{Schematic drawing of the repeater QKD protocol between two clients $A$ and $B$. The untrusted third party has prepared a long distance entangled pair between nodes $S_1$ and $S_2$. Client $A$ ($B$) remotely prepares a qubit of the form $\ket{+_\theta}$ on node $S_A$ ($S_B$). Once remote state preparation succeeds for client $A$ ($B$), a local Bell state measurement is performed between nodes $S_A$ and $S_1$ ($S_A$ and $S_2$).}
    \label{fig:QKD_drawing}
\end{figure}

RSP finds impactful applications across multiple domains within a quantum network, for example in blind quantum computing protocols (BQC). Here, in addition to BQC, we show that our RSP protocols in combination with a repeater chain can produce perfectly correlated bits across long distances (see Fig.\ref{fig:QKD_drawing}) and we discuss the possibility for our RSP schemes to be part of a quantum key distribution (QKD) protocol. 

{\it RSP for long distance communication.-} RSP enables long distance communication over a repeater chain \cite{briegel1998quantum}. For a schematic drawing see Figure \ref{fig:QKD_drawing}, where we have two clients wanting to establish a key with each other, and a repeater chain between them. Here we give the intuition behind how our setup works, and defer the calculations to the App. \ref{app:Equivalence}. The clients $A$ and $B$ remotely prepare a qubit of the form $\ket{+_\theta}$ on the first nodes of the repeater chain: $S_A$ and $S_B$. Once $A$ and $B$ have both succeeded in remotely preparing their qubits, shared entanglement between $S_A$ and $S_B$ can be used to swap the states to the same node, where a BSM can be performed between the two remotely prepared qubits. A and B use the reported measurement patterns as part of the protocol to align their bit values (with B flipping his bit when necessary), the security of the protocol does not rely on trusting these measurements. Instead, security is guaranteed through subsequent parameter estimation and privacy amplification steps between A and B.\\
In measurement-device-independent (MDI) QKD, the security of the protocol is independent on how the BSM is performed. Therefore, a protocol that uses RSP and teleportation across a repeater chain can be equivalent to an MDI QKD protocol that uses a simple BSM. We show in Appendix \ref{app:Equivalence} that these protocols for RSP are compatible with existing QKD security proofs.

{\it RSP for BQC.-} BQC using double-click RSP with a WCP-based client has been proven secure \cite{garnier2024composably} (which improves on \cite{dunjko2012blind} by eliminating the need for photon counting, adding verifiability and providing scaling in terms of number of samples with respect to the transmittance of the channel). A BQC security proof for SC and DSC is still an open question. A difficulty that arises here is that a common technique used in security proofs both in QKD and BQC with WCP sources is to phase randomize the WCP, effectively getting rid of the coherences within the WCP. This assumption cannot be made in the SC security proof, as we rely on the phase of the WCP to define the phase of the remotely prepared state. In DSC, however, the phase of the remotely prepared state depends on the phase difference between the first and the second click, such that the pulses can be phase randomized, as long as the randomization is equal for both pulses. This may allow extending standard security proofs to DSC, but a complete investigation of this is beyond the scope of this work.

\section{Conclusion}
We have introduced two powerful new protocols for remote state preparation using weak coherent pulses: single-click (SC) and double-single-click (DSC). In SC, the phase transferred onto the remotely prepared state is encoded in the weak coherent pulse of the client, while the photon-emission probability of the server is adjusted to balance photon arrival probabilities at the Bell state measurement station. DSC repeats this process twice and applies a CNOT gate to the resulting qubits, effectively canceling out random phase fluctuations between the pulses and eliminating the need for phase stabilization when the clicks occur faster than the phase fluctuations.\\
Our key findings are:
\begin{enumerate}
\item The SC protocol consistently achieves higher rates than DC while maintaining comparable fidelity levels. This is in contrast to entanglement generation, where obtaining a higher rate due to switching from DC to SC comes at a cost of lowered fidelity. This represents a significant advantage for practical implementations where high preparation rates are crucial.

\item For systems with low server efficiency, high target rates, or modest fidelity requirements, SC offers clear advantages over DC. This is especially relevant for quantum computing applications requiring multiple remote state preparations.

\item The DSC protocol provides a practical middle ground - while delivering more modest rate improvements over DC, it can eliminate the need for phase stabilization when the system can perform attempts fast enough that the phase remains stable between consecutive successes.
\end{enumerate}

The choice between these protocols depends on specific experimental constraints:
\begin{itemize}
\item Use SC when maximum rate is paramount and phase stabilization is feasible
\item Consider DSC when phase stabilization is challenging but system speeds allow for sufficiently low phase noise between consecutive clicks
\item Stick with DC for applications requiring maximum fidelity for a single qubit or when providing a low phase noise environment for SC and DSC is not feasible
\end{itemize}

We've demonstrated how these protocols can be applied to quantum key distribution over repeater chains using existing measurement-device-independent protocols. The path forward for blind quantum computing applications appears promising, particularly given recent security proofs for DC-based protocols.

\section*{Acknowledgements}
The authors would like to thank Arian Stolk, Harold Ollivier and Maxime Garnier for useful discussions, and Bethany Davies for useful comments on the manuscript.\\
This project has received funding from the European Union’s Horizon Europe research and innovation programme under grant agreement No. 101102140.\\
ERH and ASS acknowledge the support of Danmarks Grundforskningsfond (DNRF Grant No. 139, Hy-Q Center for Hybrid Quantum Networks).

\bibliographystyle{unsrt}
\bibliography{bib}

\begin{thebibliography}{10}

\bibitem{bennett2001remote}
Charles~H Bennett, David~P DiVincenzo, Peter~W Shor, John~A Smolin, Barbara~M
  Terhal, and William~K Wootters.
\newblock Remote state preparation.
\newblock {\em Physical Review Letters}, 87(7):077902, 2001.

\bibitem{broadbent2009universal}
Anne Broadbent, Joseph Fitzsimons, and Elham Kashefi.
\newblock Universal blind quantum computation.
\newblock In {\em 2009 50th Annual IEEE Symposium on Foundations of Computer
  Science}, pages 517--526. IEEE, 2009.

\bibitem{schmidt2020memory}
Frank Schmidt and Peter van Loock.
\newblock Memory-assisted long-distance phase-matching quantum key
  distribution.
\newblock {\em Physical Review A}, 102(4):042614, 2020.

\bibitem{langenfeld2021quantum}
Stefan Langenfeld, Philip Thomas, Olivier Morin, and Gerhard Rempe.
\newblock Quantum repeater node demonstrating unconditionally secure key
  distribution.
\newblock {\em Physical review letters}, 126(23):230506, 2021.

\bibitem{morimae2013blind}
Tomoyuki Morimae and Keisuke Fujii.
\newblock Blind quantum computation protocol in which alice only makes
  measurements.
\newblock {\em Physical Review A}, 87(5):050301, 2013.

\bibitem{fitzsimons2017private}
Joseph~F Fitzsimons.
\newblock Private quantum computation: an introduction to blind quantum
  computing and related protocols.
\newblock {\em npj Quantum Information}, 3(1):23, 2017.

\bibitem{vanDam2024}
J~van Dam, G~Avis, Tz~B Propp, F~Ferreira~da Silva, J~A Slater, T~E Northup,
  and S~Wehner.
\newblock Hardware requirements for trapped-ion-based verifiable blind quantum
  computing with a measurement-only client.
\newblock {\em Quantum Science and Technology}, 9(4):045031, August 2024.

\bibitem{ekert1991quantum}
Artur~K Ekert.
\newblock Quantum cryptography based on bell’s theorem.
\newblock {\em Physical review letters}, 67(6):661, 1991.

\bibitem{bennett1992quantum}
Charles~H Bennett, Gilles Brassard, and N~David Mermin.
\newblock Quantum cryptography without bell’s theorem.
\newblock {\em Physical review letters}, 68(5):557, 1992.

\bibitem{beveratos2002single}
Alexios Beveratos, Rosa Brouri, Thierry Gacoin, Andr{\'e} Villing,
  Jean-Philippe Poizat, and Philippe Grangier.
\newblock Single photon quantum cryptography.
\newblock {\em Physical review letters}, 89(18):187901, 2002.

\bibitem{garnier2024composably}
Maxime Garnier, Dominik Leichtle, Luka Music, and Harold Ollivier.
\newblock Composably secure delegated quantum computation with weak coherent
  pulses.
\newblock In {\em 2024 International Conference on Quantum Communications,
  Networking, and Computing (QCNC)}, pages 221--225. IEEE, 2024.

\bibitem{li2021blind}
Qin Li, Chengdong Liu, Yu~Peng, Fang Yu, and Cai Zhang.
\newblock Blind quantum computation where a user only performs single-qubit
  gates.
\newblock {\em Optics \& Laser Technology}, 142:107190, 2021.

\bibitem{morimae2013secure}
Tomoyuki Morimae and Keisuke Fujii.
\newblock Secure entanglement distillation for double-server blind quantum
  computation.
\newblock {\em Physical Review Letters}, 111:020502, Jul 2013.

\bibitem{sheng2015deterministic}
Yu-Bo Sheng and Lan Zhou.
\newblock Deterministic entanglement distillation for secure double-server
  blind quantum computation.
\newblock {\em Scientific reports}, 5(1):7815, 2015.

\bibitem{li2014triple}
Qin Li, Wai~Hong Chan, Chunhui Wu, and Zhonghua Wen.
\newblock Triple-server blind quantum computation using entanglement swapping.
\newblock {\em Physical Review A}, 89(4):040302, 2014.

\bibitem{quan2023verifiable}
Junyu Quan, Qin Li, and Lvzhou Li.
\newblock Verifiable blind quantum computation with identity authentication for
  multi-type clients.
\newblock {\em IEEE Transactions on Information Forensics and Security}, 2023.

\bibitem{hwang2003quantum}
Won-Young Hwang.
\newblock Quantum key distribution with high loss: toward global secure
  communication.
\newblock {\em Physical review letters}, 91(5):057901, 2003.

\bibitem{lo2005decoy}
Hoi-Kwong Lo, Xiongfeng Ma, and Kai Chen.
\newblock Decoy state quantum key distribution.
\newblock {\em Phys. Rev. Lett.}, 94:230504, Jun 2005.

\bibitem{wang2005decoy}
Xiang-Bin Wang.
\newblock Beating the photon-number-splitting attack in practical quantum
  cryptography.
\newblock {\em Phys. Rev. Lett.}, 94:230503, Jun 2005.

\bibitem{lucamarini2018overcoming}
Marco Lucamarini, Zhiliang~L Yuan, James~F Dynes, and Andrew~J Shields.
\newblock Overcoming the rate--distance limit of quantum key distribution
  without quantum repeaters.
\newblock {\em Nature}, 557(7705):400--403, 2018.

\bibitem{liao2018satellite}
Sheng-Kai Liao, Wen-Qi Cai, Johannes Handsteiner, Bo~Liu, Juan Yin, Liang
  Zhang, Dominik Rauch, Matthias Fink, Ji-Gang Ren, Wei-Yue Liu, et~al.
\newblock Satellite-relayed intercontinental quantum network.
\newblock {\em Physical review letters}, 120(3):030501, 2018.

\bibitem{scarani2009security}
Valerio Scarani, Helle Bechmann-Pasquinucci, Nicolas~J Cerf, Miloslav
  Du{\v{s}}ek, Norbert L{\"u}tkenhaus, and Momtchil Peev.
\newblock The security of practical quantum key distribution.
\newblock {\em Reviews of modern physics}, 81(3):1301--1350, 2009.

\bibitem{dunjko2012blind}
Vedran Dunjko, Elham Kashefi, and Anthony Leverrier.
\newblock Blind quantum computing with weak coherent pulses.
\newblock {\em Physical review letters}, 108(20):200502, 2012.

\bibitem{barrett2005efficient}
Sean~D Barrett and Pieter Kok.
\newblock Efficient high-fidelity quantum computation using matter qubits and
  linear optics.
\newblock {\em Physical Review A—Atomic, Molecular, and Optical Physics},
  71(6):060310, 2005.

\bibitem{simon2003robust}
Christoph Simon and William~TM Irvine.
\newblock Robust long-distance entanglement and a loophole-free bell test with
  ions and photons.
\newblock {\em Physical review letters}, 91(11):110405, 2003.

\bibitem{campbell2008measurement}
Earl~T Campbell and Simon~C Benjamin.
\newblock Measurement-based entanglement under conditions of extreme photon
  loss.
\newblock {\em Physical review letters}, 101(13):130502, 2008.

\bibitem{humphreys2018deterministic}
Peter~C Humphreys, Norbert Kalb, Jaco~PJ Morits, Raymond~N Schouten, Raymond~FL
  Vermeulen, Daniel~J Twitchen, Matthew Markham, and Ronald Hanson.
\newblock Deterministic delivery of remote entanglement on a quantum network.
\newblock {\em Nature}, 558(7709):268--273, 2018.

\bibitem{jiang2019remote}
Yang-Fan Jiang, Kejin Wei, Liang Huang, Ke~Xu, Qi-Chao Sun, Yu-Zhe Zhang,
  Weijun Zhang, Hao Li, Lixing You, Zhen Wang, et~al.
\newblock Remote blind state preparation with weak coherent pulses in the
  field.
\newblock {\em Physical Review Letters}, 123(10):100503, 2019.

\bibitem{patentsc}
Janice van Dam, Emil Hellebek, Tzula Propp, Junior Gonzales-Ureta, Anders
  S{\o}rensen, and Stephanie Wehner.
\newblock Single-click protocol for remote state preparation with a weak
  coherent pulse source.
\newblock Dutch Patent Application, 2025.
\newblock Patent pending, Application filed July 10, 2025.

\bibitem{patentdsc}
Janice van Dam, Emil Hellebek, Tzula Propp, Junior Gonzales-Ureta, Anders
  S{\o}rensen, and Stephanie Wehner.
\newblock Double-single-click protocol for remote state preparation with a weak
  coherent pulse source.
\newblock Dutch Patent Application, 2025.
\newblock Patent pending, Application filed July 10, 2025.

\bibitem{ma2012phasemdi}
Xiongfeng Ma and Mohsen Razavi.
\newblock Alternative schemes for measurement-device-independent quantum key
  distribution.
\newblock {\em Phys. Rev. A}, 86:062319, Dec 2012.

\bibitem{kapourniotis2023asymmetric}
Theodoros Kapourniotis, Elham Kashefi, Dominik Leichtle, Luka Music, and Harold
  Ollivier.
\newblock Asymmetric quantum secure multi-party computation with weak clients
  against dishonest majority.
\newblock {\em arXiv preprint arXiv:2303.08865}, 2023.

\bibitem{Note1}
This is because, the photon coming from the client leaves the server qubit in
  $\ket {0}$, because the server did not emit a photon and thus was not in the
  bright state, and the photon coming from the server leaves the qubit in $\ket
  {1}$, equal probabilities give a state with equal weights $\ket {0}$ and
  $\ket {1}$, hence on the equator of the Bloch sphere.

\bibitem{stolk2024extendable}
AJ~Stolk, JJB Biemond, KL~van~der Enden, L~van Dooren, EJ~van Zwet, and
  R~Hanson.
\newblock Extendable optical phase synchronization of remote and independent
  quantum network nodes over deployed fibers.
\newblock {\em arXiv preprint arXiv:2408.12464}, 2024.

\bibitem{czegledi2016polarization}
Cristian~B Czegledi, Magnus Karlsson, Erik Agrell, and Pontus Johannisson.
\newblock Polarization drift channel model for coherent fibre-optic systems.
\newblock {\em Scientific reports}, 6(1):21217, 2016.

\bibitem{schupp2021interface}
Josef Schupp, Vojtech Krcmarsky, Viktor Krutyanskiy, Martin Meraner, Tracy~E
  Northup, and Ben~P Lanyon.
\newblock Interface between trapped-ion qubits and traveling photons with
  close-to-optimal efficiency.
\newblock {\em PRX quantum}, 2(2):020331, 2021.

\bibitem{krutyanskiy2019light}
Viktor Krutyanskiy, Martin Meraner, Josef Schupp, Vojtech Krcmarsky, Helene
  Hainzer, and Ben~P Lanyon.
\newblock Light-matter entanglement over 50 km of optical fibre.
\newblock {\em npj Quantum Information}, 5(1):72, 2019.

\bibitem{davies2024tools}
Bethany Davies, Thomas Beauchamp, Gayane Vardoyan, and Stephanie Wehner.
\newblock Tools for the analysis of quantum protocols requiring state
  generation within a time window.
\newblock {\em IEEE Transactions on Quantum Engineering}, 2024.

\bibitem{briegel1998quantum}
H-J Briegel, Wolfgang D{\"u}r, Juan~I Cirac, and Peter Zoller.
\newblock Quantum repeaters: the role of imperfect local operations in quantum
  communication.
\newblock {\em Physical Review Letters}, 81(26):5932, 1998.

\bibitem{curty2019simple}
Marcos Curty, Koji Azuma, and Hoi-Kwong Lo.
\newblock Simple security proof of twin-field type quantum key distribution
  protocol.
\newblock {\em npj Quantum Information}, 5(1):64, 2019.

\bibitem{lo2012mdiqkd}
Hoi-Kwong Lo, Marcos Curty, and Bing Qi.
\newblock Measurement-device-independent quantum key distribution.
\newblock {\em Phys. Rev. Lett.}, 108:130503, Mar 2012.

\end{thebibliography}
\newpage
\appendix
\onecolumngrid
\section{Derivation of rate and fidelity formulas}\label{app:derivation}
Here, we will go through the calculations for finding the success probability and fidelity of a single remotely prepared qubit state using non-photon number resolving detectors. In these calculations, the target state is given by $\ket{+_\theta}$, for some angle $\theta$. We will give detailed calculations for the SC protocol. As the derivation for the other protocols are very similar, we will just state the results for the DC and DSC protocols for brevity.\\
We consider losses, which can be due to server inefficiencies, detector inefficiencies, fiber losses, losses due to frequency conversion or any other. We assume no other imperfections aside from the WCP source not being a perfect single-photon source and the phase noise discussed in \ref{subapp:phasenoise}. i.e., we assume perfect gates, the server emits with perfect fidelity and no decoherence.

\subsection{Double-click}
For double-click, the client sends out a WCP with displacement $\alpha$, lets it fall onto a beamsplitter and includes a phase shift for one of the arms. This can be a polarizing beamsplitter in the case of polarization encoding, or a regular beamsplitter with delay line in the case of time-bin encoding. The server will emit a photon entangled with the state of the qubit. We set $\xi=1/\sqrt{2}$ to maximize the fidelity; this means the server and photon qubits form a Bell state. The input states are thus given by 
\begin{align}
    \ket{\psi_c} &=e^{-\abs{\alpha}^2/2} \sum_{n,m} \frac{(\abs{\alpha}/\sqrt{2})^{n+m}}{\sqrt{n!m!}}e^{in\theta} \ket{nm},\qquad
    &\ket{\psi_s} = \sqrt{\frac{1}{2}} (\ket{\emptyset 1, 0}+ \ket{1\emptyset, 1}),
\end{align}
We now introduce losses for both the server and client photons. Furthermore, we subject the light in both the early and late time bins (assuming time-bin encoding) to a BSM. The operation will succeed if a photon is detected in both time bins. The density matrix and fidelity of the remotely prepared qubit and success probability of the protocol are
\begin{align}\label{appeq:rho_DC}
    \rho_\text{DC} &= \frac{1}{2}\qty{\mathbb{I}+\frac{(e^{-i\theta}\ketbra{0}{1}+\text{h.c.})\eta_s\eta_c\abs{\alpha}^2/8}{\pqty{1-e^{-\eta_c\abs{\alpha}^2/4}}\bqty{\frac{\eta_s}{2}\pqty{1+\frac{\eta_c\abs{\alpha}^2}{4}}+\pqty{1-e^{-\eta_c\abs{\alpha}^2/4}}\pqty{1-\eta_s}}}},\\\label{appeq:F_DC}
    F_\text{DC} &= \frac{1}{2}\qty{
    1+\frac{\eta_s\eta_c\abs{\alpha}^2/8}{\pqty{1-e^{-\eta_c\abs{\alpha}^2/4}}\bqty{\frac{\eta_s}{2}\pqty{1+\frac{\eta_c\abs{\alpha}^2}{4}}+\pqty{1-e^{-\eta_c\abs{\alpha}^2/4}}\pqty{1-\eta_s}}}}\sim 1-\frac{\eta_c}{\eta_s}\frac{4-3\eta_s}{16}\abs{\alpha}^2,\\\label{appeq:P_DC}
    P_\text{DC} &= 4e^{-\eta_c\abs{\alpha}^2/2}\pqty{1-e^{-\eta_c\abs{\alpha}^2/4}}\bqty{\frac{\eta_s}{2}\pqty{1+\frac{\eta_c\abs{\alpha}^2}{4}}+\pqty{1-e^{-\eta_c\abs{\alpha}^2/4}}\pqty{1-\eta_s}}\sim \frac{\eta_c\eta_s}{2} \abs{\alpha}^2.
\end{align}

\subsection{Single-click}
The client sends out a coherent state with displacement $\alpha=\abs{\alpha}e^{-i\theta}$. The server sends out a single photon when its memory qubit is in the bright state, denoted $\ket{1}$ (and no photon otherwise). The probability of the server being in the bright state is dependent on the bright state parameter $\xi$. Thus, we start with the following states
\begin{align}\label{appeq:SCinputs}
    \ket{\psi_c} &= \ket{\alpha}=e^{-\abs{\alpha}^2/2} \sum_n \frac{\alpha^n}{\sqrt{n!}} \ket{n},\qquad
    &\ket{\psi_s} = \sqrt{1-\xi^2} \ket{\emptyset 0}+\xi \ket{11}.
\end{align}
We now introduce losses $1-\eta_{c}$ and $1-\eta_{s}$ on the client and server states. This transforms the client state to $\ket{\sqrt{\eta_c}\alpha}$, by effectively rescaling the mean photon number for the WCP, and the server is represented by the density-matrix
\begin{align}\label{appeq:server_dm}
    \rho_S = \,&\left(\sqrt{1-\xi^2}\ket{\emptyset 0}+ \sqrt{\eta_s}\xi\ket{11}\right)(\text{h.c.})+\xi^2(1-\eta_s) \ketbra{\emptyset1}{\emptyset1},
\end{align}
meaning the system is given by the density matrix $\rho_\text{sys}=\rho_S\otimes\ketbra{\sqrt{\eta_c}\alpha}$.
The 50/50 beamsplitter transforms the client and server photons into a plus and minus modes with the annihilation operators $a_\pm=(a_s \pm a_c)/\sqrt{2}$, where $a_c$ ($a_s$) is the annihilation operator of the client (server) photon. A photon detector is placed in both paths, and will be referred to as plus and minus detectors, respectively.
We will assume that a signal comes from the plus detector while the minus detector stays silent, corresponding to the measurement operator $(1-\ketbra{\emptyset_+})\ketbra{\emptyset_-}$, leading to density matrix
\begin{align}\label{appeq:rho_trans}
    \rho_{\text{sys}}\to \rho_\text{SC}=\frac{\tr_\phi\qty[\qty(1-\ketbra{\emptyset_+})\ev{\rho_\text{sys}}{\emptyset_-}\qty(1-\ketbra{\emptyset_+})]}{\tr\qty[\qty(1-\ketbra{\emptyset_+})\ev{\rho_\text{sys}}{\emptyset_-}\qty(1-\ketbra{\emptyset_+})]},
\end{align}
where $\tr_\phi$ is the trace over the photon subspace, $\ket{\emptyset_-}$ is the state of no photons in the minus detector, and the denominator is the probability of obtaining a click in the plus detector, $P_{\text{SC}|+}$.
The numerator is given by
\begin{align}
    \tr_\phi&\qty[\qty(1-\ketbra{\emptyset_+})\ev{\rho_\text{sys}}{\emptyset_-}\qty(1-\ketbra{\emptyset_+})]=e^{-\eta_c\abs{\alpha}^2/2}\bigg\{
    \pqty{1-e^{-\eta_c\abs{\alpha}^2/2}}\pqty{1-\xi^2}\ketbra{0}\\\nonumber
    &+\frac{\sqrt{\eta_c\eta_s(1-\xi^2)}}{2}\abs{\alpha}\xi\pqty{e^{i\theta}\ketbra{1}{0}+\text{h.c.}}+\xi^2\bqty{\frac{\eta_s}{2}\pqty{1+\frac{\eta_c\abs{\alpha}^2}{2}}+\pqty{1-\eta_s}\pqty{1-e^{-\eta_c\abs{\alpha}^2/2}}}\ketbra{1}
    \bigg\}.
\end{align}
We next find the denominator, $P_{\text{SC}|+}$ 
\begin{align}
    P_{\text{SC}|+}=e^{-\eta_c\abs{\alpha}^2/2}\bqty{
    \pqty{1-e^{-\eta_c\abs{\alpha}^2/2}}\pqty{1-\eta_s\xi^2}+\xi^2\frac{\eta_s}{2}\pqty{1+\frac{\eta_c\abs{\alpha}^2}{2}}
    },
\end{align}
where the success probability is $P_\text{SC} = 2P_{\text{SC}|+}$, as $P_{\text{SC}|+}=P_{\text{SC}|-}$. The density matrix after the heralding click is then
\begin{align}\label{appeq:rhosc}
    \rho_\text{SC} = &\frac{
    \pqty{1-e^{-\eta_c\abs{\alpha}^2/2}}\pqty{1-\xi^2}\ketbra{0}+\frac{\sqrt{\eta_c\eta_s(1-\xi^2)}}{2}\abs{\alpha}\xi\pqty{e^{i\theta}\ketbra{1}{0}+\text{h.c.}}}{\pqty{1-e^{-\eta_c\abs{\alpha}^2/2}}\pqty{1-\eta_s\xi^2}+\xi^2\frac{\eta_s}{2}\pqty{1+\frac{\eta_c\abs{\alpha}^2}{2}}}\\\nonumber
    &+\frac{\xi^2\bqty{\frac{\eta_s}{2}\pqty{1+\frac{\eta_c\abs{\alpha}^2}{2}}+\pqty{1-\eta_s}\pqty{1-e^{-\eta_c\abs{\alpha}^2/2}}}\ketbra{1}}{\pqty{1-e^{-\eta_c\abs{\alpha}^2/2}}\pqty{1-\eta_s\xi^2}+\xi^2\frac{\eta_s}{2}\pqty{1+\frac{\eta_c\abs{\alpha}^2}{2}}}.
\end{align}
We can now find the fidelity with respect to our target state $\ket{+_\theta}$ as $F=|\bra{+_\theta}\rho_\text{SC}\ket{+_\theta}|$ and find
\begin{align}
    F= \frac{1}{2}\bqty{1+\frac{\sqrt{\eta_c\eta_s(1-\xi^2)}\abs{\alpha}\xi}{\pqty{1-e^{-\eta_c\abs{\alpha}^2/2}}\pqty{1-\eta_s\xi^2}+\xi^2\frac{\eta_s}{2}\pqty{1+\frac{\eta_c\abs{\alpha}^2}{2}}}}.
\end{align}
This fidelity can be maximized with respect to the bright state parameter $\xi$, leading to
\begin{align}
    \xi_\text{SC}^2=\frac{2 \pqty{1-e^{-\eta_c\abs{\alpha}^2/2}}}{
    4\pqty{1-e^{-\eta_c\abs{\alpha}^2/2}}+\eta_s\pqty{2 e^{-\eta_c\abs{\alpha}^2/2}+\eta_c\abs{\alpha}^2-1
    }}\sim \frac{\eta_c}{\eta_s}\abs{\alpha}^2.
\end{align}
Here, the last expression is the expansion up to first order in the mean photon number from the client, as we will want to be in a regime where the client sends out a weak pulse. We continue the calculations with the exact expression, not the first order expansion. The fidelity becomes
\begin{align}\label{eq:Fscapp}
    F_\text{SC}=\frac{1}{2}\qty{1+\sqrt{\frac{\eta_c\eta_s\abs{\alpha}^2/2}{\pqty{1-e^{-\eta_c\abs{\alpha}^2/2}}\bqty{2\pqty{1-\eta_s}\pqty{1-e^{-\eta_c\abs{\alpha}^2/2}}+\eta_s\pqty{1+\eta_c\abs{\alpha}^2/2}}}}}\sim 1-\frac{\eta_c(4-3\eta_s)}{16\eta_s}\abs{\alpha}^2.
\end{align}
We similarly substitute the expression for the optimized bright state parameter in the success probability to find
\begin{align}
    P_\text{SC} = 2e^{-\eta_c\abs{\alpha}^2/2}\pqty{1-e^{-\eta_c\abs{\alpha}^2/2}}\bqty{1+\frac{\eta_s\pqty{2e^{-\eta_c\abs{\alpha}^2/2}-1+\eta_c\abs{\alpha}^2/2}}{\eta_s\pqty{2e^{-\eta_c\abs{\alpha}^2/2}-1+\eta_c\abs{\alpha}^2}+4\pqty{1-e^{-\eta_c\abs{\alpha}^2/2}}}}\sim 2\eta_c\abs{\alpha}^2.
\end{align}
From this, the rate can be calculated as the success probability multiplied by the time per attempt $\tau$. We use the `dimensionless rate' $R\tau$ in our analysis, as this allows us to compare it to the DSC protocol, where one does not speak of a success probability per attempt (as we need two successes). 

\subsection{Double-single-click}
For double-single-click, the single-click protocol is performed twice. The state after the two successful clicks is $\rho_\text{sys} = \rho_\text{SC}\otimes\rho_\text{SC}$, with $\rho_\text{SC}$ as in Equation \ref{appeq:rhosc}. Then, a controlled-NOT (CNOT) gate is performed on the two prepared qubits, followed by a measurement on the target qubit. The state is only accepted if this qubit is in the bright state $\ket{1}$. As the CNOT gate and measurement ensures that the two qubits were measured in different states, the fidelity is no longer dependent on the bright state parameter. The density matrix and the fidelity of the remotely prepared state are 
\begin{align}\label{appeq:rho_DSC}
    \rho_\text{DSC} &= \frac{1}{2}\qty{\mathbb{I}+\frac{(e^{-i\theta}\ketbra{0}{1}+\text{h.c.})\eta_s\eta_c\abs{\alpha}^2/4}{\pqty{1-e^{-\frac{\eta_c\abs{\alpha}^2}{2}}}\bqty{\frac{\eta_s}{2}\pqty{1+\frac{\eta_c\abs{\alpha}^2}{2}}+\pqty{1-\eta_s}\pqty{1-e^{-\frac{\eta_c\abs{\alpha}^2}{2}}}}
    }},\\\label{appeq:F_DSC}
    F_\text{DSC} &= \frac{1} {2}\qty{1+\frac{\eta_s\eta_c\abs{\alpha}^2/4}{\pqty{1-e^{-\frac{\eta_c\abs{\alpha}^2}{2}}}\bqty{\frac{\eta_s}{2}\pqty{1+\frac{\eta_c\abs{\alpha}^2}{2}}+\pqty{1-\eta_s}\pqty{1-e^{-\frac{\eta_c\abs{\alpha}^2}{2}}}}
    }}\sim 1-\frac{\eta_c}{\eta_s}\frac{4-3\eta_s}{8}\abs{\alpha}^2.
\end{align}
It does not make sense to talk of a success probability of the process, as we will assume that we store the first qubit, while we wait for the second qubit to be generated. We will instead talk in terms of the rate, assuming a fixed time for each attempt. We will consider that we prepare the two qubits on the server in parallel, meaning, the dimensionless rate will be
\begin{align}
    R_\text{DSC}\tau = \pqty{\frac{1}{2p_\text{SC}-p_\text{SC}^2}+\frac{1}{p_\text{SC}}}^{-1}P_\text{CNOT}\sim \frac{2}{3} p_\text{SC}P_\text{CNOT},
\end{align}
where $P_\text{SC}$ is the single click success probability, with displacement $\alpha/\sqrt{2}$; the first (second) term in the parentheses corresponds to the time it takes to prepare the first (second) qubit and $P_\text{CNOT}$ is the probability that the measurement after the CNOT gate yields the correct outcome. As $P_\text{SC}$ is small, we will use the approximated version for the success probability. We choose the bright state parameter which maximizes the success probability, yielding the expression
\begin{align}
    \label{appeq:P_DSC}
    R_\text{DSC}\tau =&\, \frac{8\left(1-e^{-\frac{\eta_c|\alpha|^2}{2}}\right) \bqty{\eta _s \pqty{3+\eta_c|\alpha|^2}+4\pqty{1-e^{-\frac{|\alpha|^2 \eta _c}{2}}} \pqty{1-\eta _s}}}{3\eta _s^2e^{\frac{\eta_c\abs{\alpha}^2}{2}} \bqty{3+\eta_c|\alpha|^2-4\pqty{1-e^{-\frac{\eta_c|\alpha |^2}{2}}}}^2}
   \bigg\{8-\eta _s \left(1-2\eta_c|\alpha|^2\right)\\&\nonumber\quad-4
   \sqrt{\left(1-e^{\frac{-\eta_c| \alpha | ^2}{4}}\right) \bqty{\eta _s \left(3+\eta_c|\alpha|^2\right)+4\pqty{1-e^{\frac{-\eta_c| \alpha | ^2}{2}}}
   \left(1-\eta _s\right)}}-4e^{-\frac{\eta_c|\alpha |^2}{2}} \left(2-\eta _s\right)\bigg\}\sim 
   \frac{4}{3}\eta_c\abs{\alpha}^2.
\end{align}

\subsection{Single-click with photon number resolving detectors}\label{subapp:sc_pnr}
If we assume photon number resolving detectors, for single click the transformation now reads
\begin{align}\label{eq:rho_trans}
    \rho_{\text{sys}}\to \rho_\text{SC}=\frac{\ev{\rho_\text{sys}}{1_+}}{\tr[\ev{\rho_\text{sys}}{1_+}]},
\end{align}
which leads to the probability of success of obtaining a click in the "plus"-mode is
\begin{align}
    p_{\text{SC}|+} = \frac{e^{-\eta_c\abs{\alpha}^2}}{2}\bqty{\eta_c\abs{\alpha}^2\pqty{1-\eta_s\xi^2}+\eta_s\xi^2}.
\end{align}
The state after the heralding becomes
\begin{align}
    \rho_\text{SC} = \frac{(1-\xi^2)\eta_c\abs{\alpha}^2+\eta_s\xi^2}{\eta_c\abs{\alpha}^2\pqty{1-\eta_s\xi^2}+\eta_s\xi^2}\ketbra{\xi} + \frac{\eta_c\abs{\alpha}^2\xi^2(1-\eta_s)}{\eta_c\abs{\alpha}^2\pqty{1-\eta_s\xi^2}+\eta_s\xi^2}\ketbra{1}.
\end{align}
To obtain the ideal state we find the same relation between $\xi$ and $\abs{\alpha}$ as the scheme without photon number resolving detectors. The total success probability under this condition becomes
\begin{align}
    p_\text{SC} = e^{-\eta_c\abs{\alpha}^2}\bqty{\eta_c\abs{\alpha}^2\pqty{1-\frac{\eta_s\eta_c\abs{\alpha}^2}{\eta_s+\eta_c\abs{\alpha}^2}}+\frac{\eta_s\eta_c\abs{\alpha}^2}{\eta_s+\eta_c\abs{\alpha}^2}}\sim 2\eta_c\abs{\alpha}^2
\end{align}
The density matrix then becomes
\begin{align}
    \rho_\text{SC} = \frac{2\eta_s}{2\eta_s+\eta_c\abs{\alpha}^2\pqty{1-\eta_s}}\ketbra{\psi_\theta} + \frac{\eta_c\abs{\alpha}^2\qty(1-\eta_s)}{2\eta_s+\eta_c\abs{\alpha}^2\pqty{1-\eta_s}}\ketbra{1}.
\end{align}
Yielding the fidelity 
\begin{align}
    F_\text{SC} = \frac{2\eta_s}{2\eta_s+\eta_c\abs{\alpha}^2\pqty{1-\eta_s}}+\frac{\eta_c\abs{\alpha}^2\qty(1-\eta_s)/2}{2\eta_s+\eta_c\abs{\alpha}^2\pqty{1-\eta_s}}\sim 1-\frac{\eta_c\abs{\alpha}^2}{4\eta_s}\pqty{1-\eta_s}.
\end{align}
Thus the improvement of using number resolving detectors over non-photon number resolving detectors is the factor of $1-\eta_s$ on the first order term in the intensity of the laser, and we find perfect fidelity if there are no losses on the server side.

\subsection{Phase noise}\label{subapp:phasenoise}
Phase noise in this system arises from different underlying causes in the three protocols. For SC the phase of the coherent state $\theta$ will change as it travels to the BSM. We will thus transform the coherent state to the density matrix
\begin{align}
    \ketbra{\abs{\alpha}e^{-i\theta}}\to \int d\delta_\theta p_{\delta_\theta} \ketbra{\abs{\alpha}e^{-i(\theta+\delta_\theta)}}.
\end{align}
where $\delta_\theta$ is the added phase noise, and $p_{\delta_\theta}$ is the distribution of the phase noise. This noise, means the remotely prepared state is of the form
\begin{align}
    \rho = a\ketbra{0}+b\ketbra{1}+\int d\delta_\theta p_{\delta_\theta}\pqty{ce^{i(\theta+\delta_\theta)}\ketbra{1}{0}+ ce^{-i(\theta+\delta_\theta)}\ketbra{0}{1}},
\end{align}
where $a,$ $b$ and $c$ are the positive real coefficients of the single click density matrix. We will assume $p_{\delta_\theta}$ to be a Gaussian distribution with variance $\sigma_{SC}^2$. In principle one should use a wrapped Gaussian distribution, but it will not change our results, as we will integrate over a periodic function. This leads us to the average fidelity
\begin{align}
    F = \frac{1+2ce^{-\sigma_{SC}^2/2}}{2}.
\end{align}
In the DSC and DC protocols, the density matrix will be transformed by the noise such that
\begin{align}
    \ketbra{\abs{\alpha}e^{-i\theta},\abs{\alpha}} \to \iint d\theta_1 d\theta_2 p_{\theta_1,\theta_2} \ketbra{\abs{\alpha}e^{-i(\theta+\theta_1)},\abs{\alpha}e^{-i\theta_2}},
\end{align}
where $\theta_1$ ($\theta_2$) is the random phase added to the first (second) coherent state. The remotely prepared state will take the form
\begin{align}
    \rho = a\ketbra{0}+b\ketbra{1}+\iint d\theta_1 d\theta_2  p_{\theta_1,\theta_2}\pqty{ce^{i(\theta+\theta_1-\theta_2)}\ketbra{1}{0}+ ce^{-i(\theta+\theta_1-\theta_2)}\ketbra{0}{1}},
\end{align}
where $a$, $b$ and $c$ can be found from Eqs. \eqref{appeq:rho_DSC} and \eqref{appeq:rho_DC} for DSC and DC, respectively. As, it is the difference between the two phase noises $\theta_1$ and $\theta_2$ which matters, we will make a change of variable to such that $\theta_1\to\theta_2+\delta_\theta$, this yields the density matrix
\begin{align}
    \rho = a\ketbra{0}+b\ketbra{1}+\int d\delta_\theta q_{\delta_\theta}\pqty{ce^{i(\theta+\delta_\theta)}\ketbra{1}{0}+ ce^{-i(\theta+\delta_\theta)}\ketbra{0}{1}},
\end{align}
where $q_{\delta_\theta}$ is marginal distribution over the variable $\delta_\theta$. For DC we can assume $q_{\delta_\theta}$ to be very close to a delta function as the two coherent state co-propagate with a different polarization for the polarization encoding, or with a very slow time delay for the time-bin encoding. For DSC, however, the phase noise depends on the time between consecutive clicks. The shorter the time between successful clicks, the narrower the probability distribution of the phase noise. Assuming $q_{\delta_\theta}$ is a Gaussian distribution with variance $\sigma_{DSC}^2$, we obtain the fidelity
\begin{align}
    F = \frac{1+2ce^{-\sigma_{DSC}^2/2}}{2}.
\end{align}

\section{Alternative value plots}\label{app:altplots}
\begin{figure*}[!ht]
\centering
    {\centering
     \subfloat[Protocol that gives the highest rate given a target fidelity of 0.97 for $\sigma_\text{DSC}=0.5$ rad.]{\includegraphics[width=0.33\textwidth]{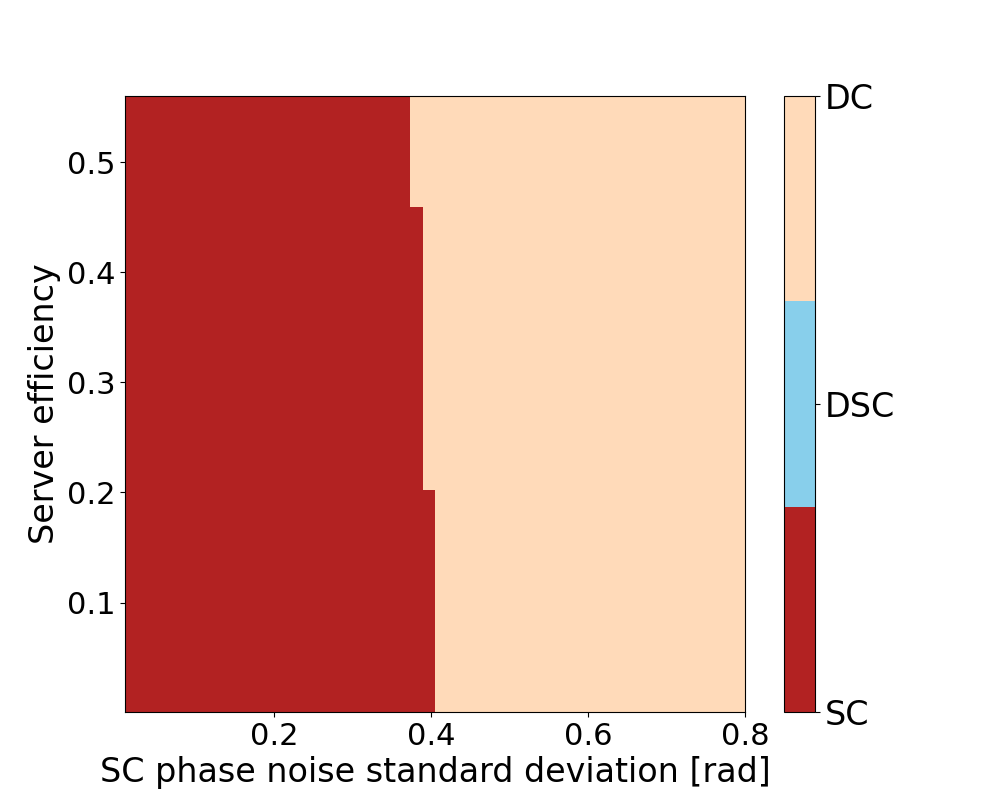}}}
    {\centering
     \subfloat[Protocol that gives the highest rate given a target fidelity of 0.97 for $\sigma_\text{DSC}=0.25$ rad.]{\includegraphics[width=0.33\textwidth]{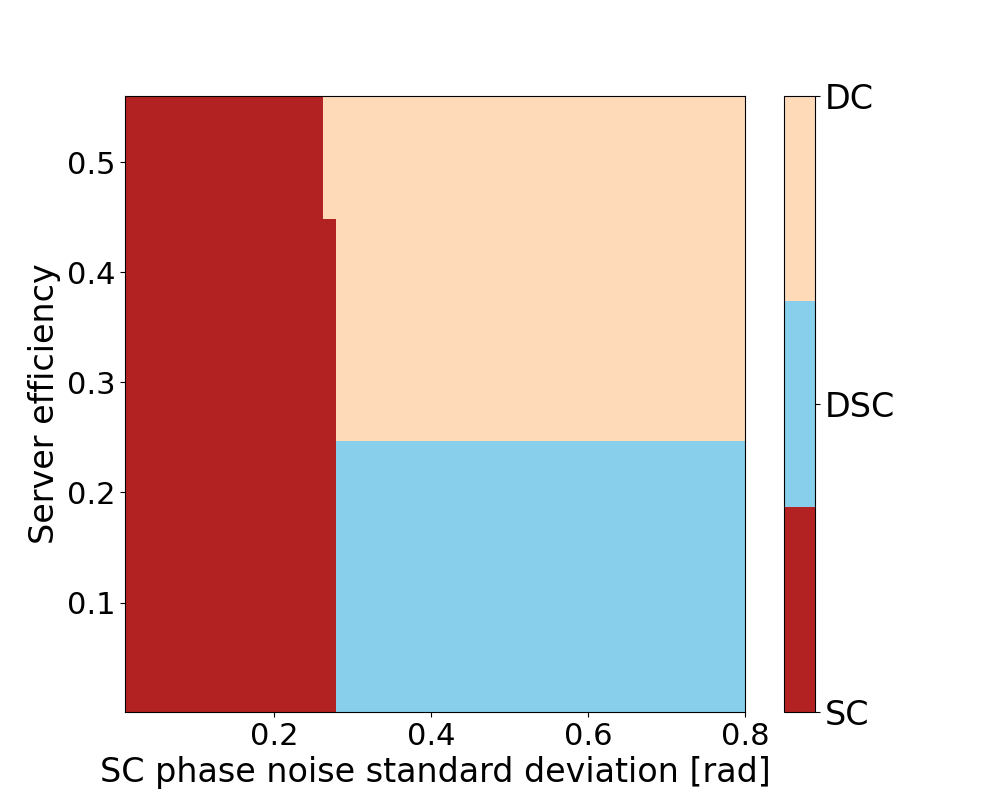}}}
    \caption{Optimal RSP protocol — DC, DSC, or SC — for a given target fidelity, across varying levels of phase noise $\sigma_\text{SC}$ and $\sigma_\text{DSC}$ (standard deviation in radians), and server efficiency. Alternative values for Figure \ref{fig:best}(a)}\label{fig:bestalta}
\end{figure*}

\begin{figure*}[!ht]
\centering
    {\centering
     \subfloat[Protocol that gives the highest fidelity given a target rate of 0.03$\tau^{-1}$ for $\sigma_\text{DSC}=0.5$ rad.]{\includegraphics[width=0.33\textwidth]{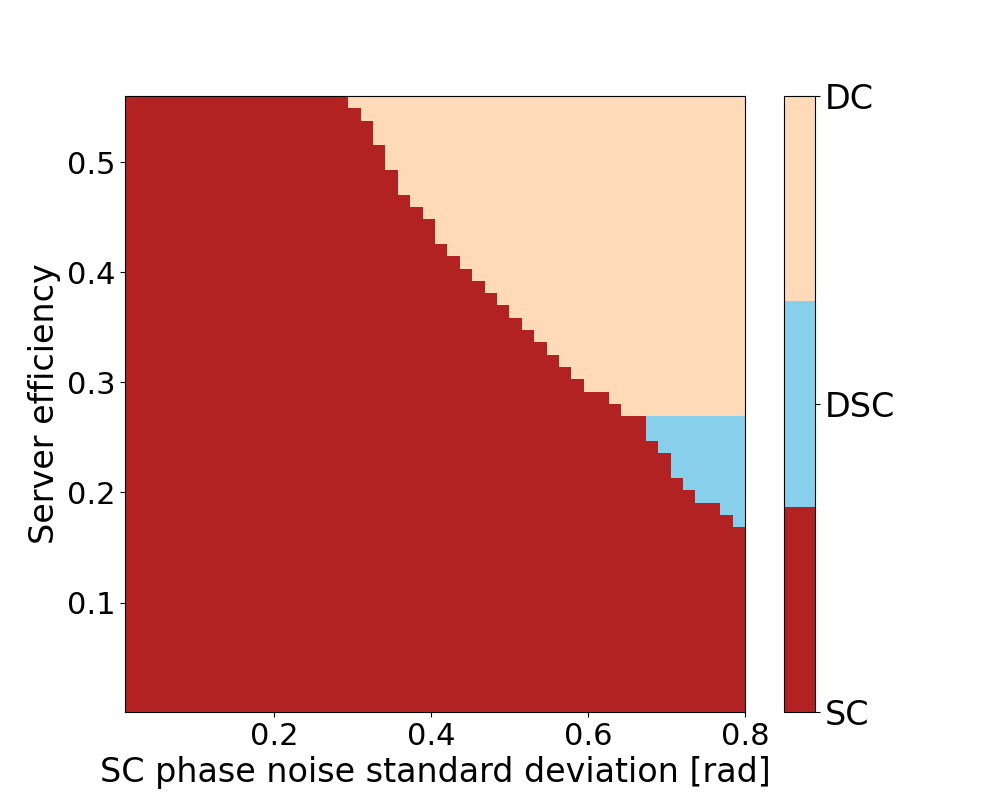}}}
    {\centering
     \subfloat[Protocol that gives the highest fidelity given a target rate of 0.01$\tau^{-1}$ for $\sigma_\text{DSC}=0.25$ rad.]{\includegraphics[width=0.33\textwidth]{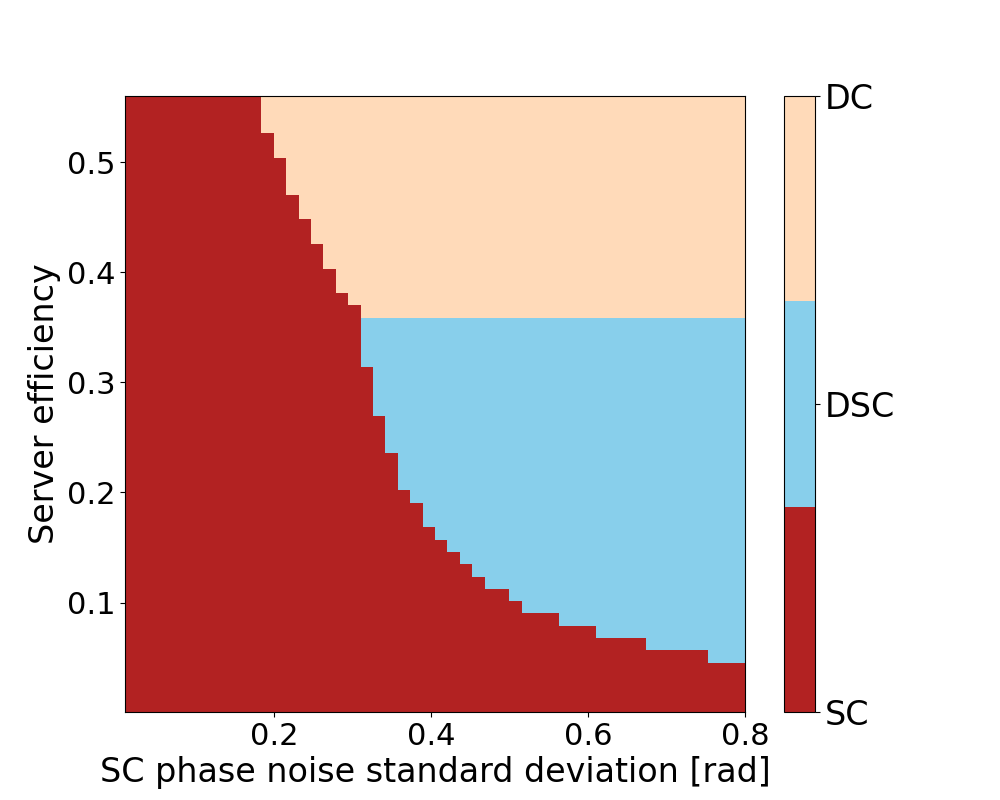}}}
    \caption{Optimal RSP protocol — DC, DSC, or SC — for a given target rate, across varying levels of phase noise $\sigma_\text{SC}$ and $\sigma_\text{DSC}$ (standard deviation in radians), and server efficiency. Alternative values for Figure \ref{fig:best}(b)}\label{fig:bestaltb}
\end{figure*}

\begin{figure*}[!ht]
\centering
    {\centering
     \subfloat[Protocol that gives the highest fidelity given a target rate of 0.01$\tau^{-1}$ for $\eta_s=0.1$.]{\includegraphics[width=0.33\textwidth]{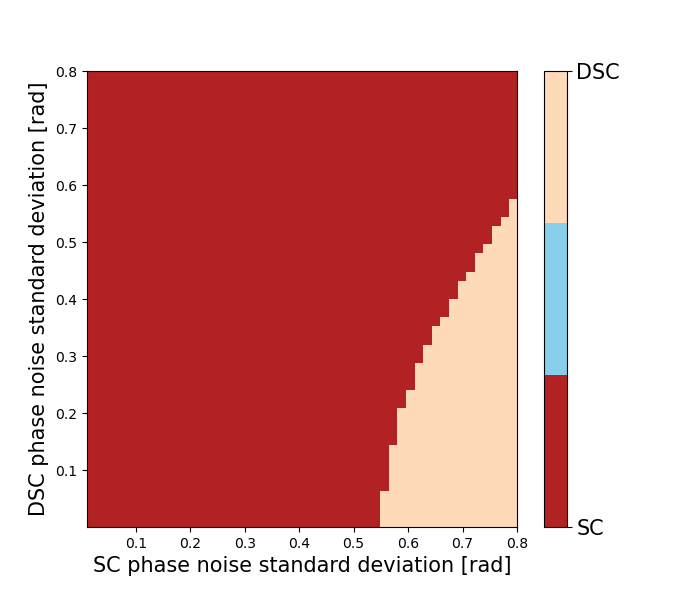}}}
    {\centering
     \subfloat[Protocol that gives the highest fidelity given a target rate of 0.01$\tau^{-1}$ for $\eta_s=0.5$.]{\includegraphics[width=0.33\textwidth]{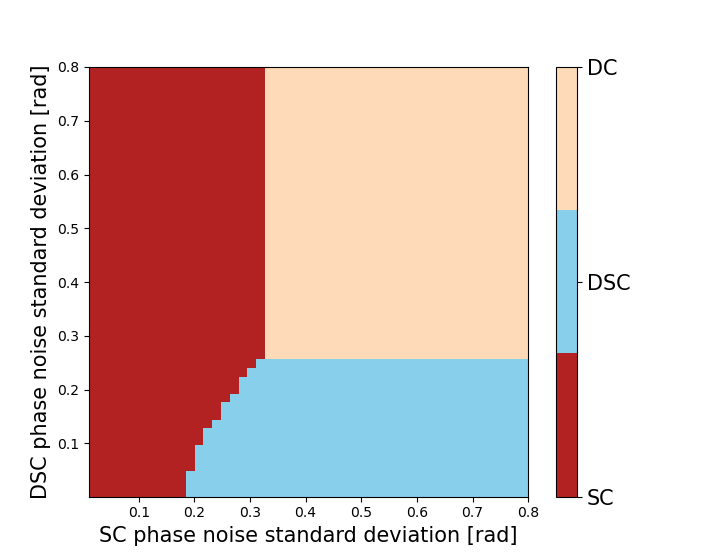}}}
    \caption{Optimal RSP protocol — DC, DSC, or SC — for a given target rate, across varying levels of phase noise $\sigma_\text{SC}$ and $\sigma_\text{DSC}$ (standard deviation in radians), and server efficiency. Alternative values for Figure \ref{fig:best}(c)}\label{fig:bestaltc}
\end{figure*}

Figures \ref{fig:bestalta}, \ref{fig:bestaltb} and \ref{fig:bestaltc} are the same as those occurring in the main text, but with alternative values to understand how the regimes changes the parameters change. Figure \ref{fig:bestalta} (a) shows that the advantage region for SC grows if the target fidelity is lower. This is because SC will be faster than DC always, and the region where SC is not shown to be faster is just because SC cannot reach the target fidelity in that area. As expected, Figure \ref{fig:bestalta} (b) shows that the advantage region for DSC grows when the phase noise for DSC is lower. Figure \ref{fig:bestaltb} shows the same behavior: more DSC phase noise gives less advantage for DSC, less DSC phase noise gives more advantage for DSC. Lastly, Figure \ref{fig:bestaltc} shows the effect of the server efficiency on the advantage regions. We see that SC is useful when the server efficiency is low (larger region for SC in \ref{fig:bestaltc} (a)) and DC regains some territory when the server is very efficient (larger region for DC in \ref{fig:bestaltc} (b)). This is also intuitive as SC comes from the idea that the photon arrival probability is low, and therefore the probability of getting two clicks is very low, this does not hold when the server is very efficient. 

\section{purified protocols for quantum key distribution}\label{app:Equivalence}
To analyze the security of QKD over a repeater chain using these RSP protocols, we convert them to a purified entanglement-based protocol. In such a purified protocol, two clients, Alice and Bob, aim to create entanglement between them. Under the assumption that Alice and Bob's devices are secure and trusted, the purified protocol is not differentiable from the real protocol by any eavesdropper on the quantum channel. The data and data processing performed by the clients is also the same for the purified protocol as for the real protocol. This allows us to analyze the security of the setup depicted in Figure \ref{fig:QKD_drawing} for performing QKD by examining the equivalent entanglement-based protocol.
Below, we go over the SC and DC purified protocol in detail to show how entanglement is formed between Alice and Bob. 

\subsection{Single-click}\label{appsubsec:SCequivalence}
Here we show that our SC protocol for remote state preparation (RSP-SC) in combination with the setup depicted in Figure\ref{fig:QKD_drawing} can be used to implement a twin-field type protocol \cite{lucamarini2018overcoming}. In particular follow the security proof given by Curty-Azuma-Lo in the so-called CAL19 protocol \cite{curty2019simple}. First we recap the steps of the CAL19 protocol, and then we illustrate how our RSP-SC protocol allows us to produce perfectly correlated bits and its compatibility with the CAL19 security proof.\\

\textbf{CAL19 - protocol}
\begin{enumerate}[label=(\roman*)]
    \item In each round of the protocol Alice and Bob can choose between two basis, the key generation basis $K$ and the test basis $T$. These bases are chosen with probability $p_K$ and $p_T$, respectively. When the basis chosen is $K$, Alice (Bob) also chooses a random bit $k_a$ ($k_b$). When the random bit is $k_a=0$ ($k_b=0$) Alice (Bob) prepares the coherent state $\ket{\alpha_a}$ ($\ket{\alpha_b}$), while for $k_a=1$ ($k_b=1$), she (he) prepares $\ket{-\alpha_a}$ ($\ket{-\alpha_b}$). When the basis chosen is $T$, Alice prepares a phase-randomized WCP $\hat{\rho}_{a, \beta_A}$ ($\hat{\rho}_{b, \beta_B}$) with a mean photon number randomly picked from a set $S=\{|\beta_i|^2\}_i$.
    
    \item Alice and Bob send their prepared optical pulses through their corresponding channels.
    
    \item In the central node $C$ both optical pulses are merged in a 50:50 beamsplitter. The result of this interference is registered by the detectors at the output ports of the beamsplitter. We denominate the detector at the first output as $D_0$, and the one on the second port as $D_{\pi}$. We use these names to stress that when the incoming coherent fields have the same phase, in the absence of any source of noise, they should produce a click in $D_0$. Although if the coherent fields have a phase difference of $\pi$, we should expect to register a click in $D_\pi$.
    
    \item After the measurement is performed, the central node $C$ announces which detectors clicked. There are four possible click patterns, but Alice and Bob will keep their data only when one click was registered, either on $D_0$ or $D_{\pi}$, the rest of cases are ignored.
    
    \item The previous steps are repeated N times such that Alice and Bob can collect enough statistics to perform parameter estimation and quantify the amount of information leaked to an eavesdropper.

    \item Finally, Alice and Bob perform classical error correction and privacy amplification to obtain a final secret key.
\end{enumerate}

There are two observations to be made regarding this protocol. First, to simplify the explanation, consider that we only post-select the clicks in $D_0$ and that we neglect sources of noise. Then, we see that all events in the $K$ basis produce perfectly correlated bits for Alice and Bob, i.e., $k_a = k_b$. This is desirable for a QKD protocol, since a protocol that in the noiseless scenario would not produce perfect correlations would lead to higher costs for error correction in the post-processing stage.

Now we detail our protocol, which is based on the scheme for RSP-SC (see \ref{subsec:scanalytics})\\

\textbf{SC RSP based protocol}
\begin{enumerate}[label=(\roman*)]
    \item A qubit in server nodes $S_A$ and $S_B$ (see Fig.\ref{fig:QKD_drawing}), which are at the border of the repeater chain, is prepared in the state $\sqrt{1-\xi^2}\ket{0} + \xi \ket{1}$, with $\xi$ the bright-state parameter of the server qubit. Repeater nodes $S_1$ and $S_2$ share entanglement of the form $\ket{\Phi^+}=(\ket{00} + \ket{11})/\sqrt{2}$.

    \item Alice and Bob both prepare the state
    \begin{align}
        \ket{\psi_{A/B}} = \frac{1}{\sqrt{2}}(\ket{\alpha,0}+\ket{-\alpha,1}),
    \end{align}
    which combines the state of the photon, with $\alpha$ is a positive real number, with the state of Alice's or Bob's register.
    
    \item The server node $S_A$ emits a photon entangled with a qubit at the node. Here, we assume perfect efficiencies and photon number resolving detectors, meaning we can perform RSP on $S_{A/B}$ with perfect fidelity (see  Appendix \ref{subapp:sc_pnr}). Then, performing a BSM between Alice and node $S_A$ leads to an entangled state between Alice's register and the node 
    \begin{align}
        \ket{\psi_{A,S_A}} = \frac{\alpha\sqrt{1-\xi^2}\ket{0}\ket{-}+\xi\ket{1}\ket{+}}{\sqrt{\alpha^2(1-\xi^2)+\xi^2}},
    \end{align}
    where $\ket{\pm}=(\ket{0}\pm\ket{1})/\sqrt{2}$. The same is repeated on Bob's side.

    \item The entanglement is propagated through the repeater chain using entanglement swapping until Alice's and Bob's registers share this state with neighboring nodes, such as $S_2$ and $S_B$. Then, a BSM is performed on $S_2$ and $S_B$ of $\ket{\psi_{A,S_2} \otimes \psi_{B,S_B}}$. With this, the state shared by Alice's and Bob's registers is projected onto (up to some corrections on Bob's side depending on the click pattern of the BSM)
    \begin{equation}
        \ket{\psi_{A,B}}=\frac{\alpha^2(1-\xi^2) \ket{00} + \xi^2 \ket{11}}{\sqrt{\alpha^4(1-\xi^2)^2 + \xi^4}}.
    \end{equation}
    This is equals the Bell state $\ket{\Phi^+}_{A,B}$ for $\xi=\alpha/\sqrt{1+\alpha^2}$. 
\end{enumerate}

With this, we see that the RSP-SC based protocol achieves the same functionality as the CAL19 protocol in the noiseless case, both of them can produce perfectly correlated bits for Alice and Bob. In order to claim that our RSP-SC based protocol can produce a secure key, we need to take into account the multi-photon contributions in the WCP used. In the CAL19 this is taken into account using the test basis together with the decoy state method \cite{lo2005decoy,wang2005decoy}. In order to include a test basis, we can simply phase-randomized the sources of Alice and Bob. And consider a set of different intensities for their pulses in order to also implement the decoy state method. A comparison of the performance of both protocol in noisy scenarios is left for future work.

\subsection{Double-click}\label{appsubsec:DCequivalence}
In this protocol, Alice and Bob send out BB84 states \cite{lo2012mdiqkd}. In the fictious entanglement-based protocol, for the $Z$ basis, Alice and Bob start with a state
\begin{equation}
    \ket{\psi_{A/B}}=\frac{1}{\sqrt{2}}(\ket{0,\alpha_a} + \ket{1,\alpha_b}) 
\end{equation}
with $\ket{\alpha_{a/b}}=e^{-\abs{\alpha}^2} \sum_{n=0}^\infty \frac{\alpha^n}{\sqrt{n!}} (a^\dagger/b^\dagger)^n\ket{0}$, where $a^\dagger$ and $b^\dagger$ are the creation operators for modes $a$ and $b$ (e.g., horizontal/vertical polarization, early/late time bin), respectively. 
For the $X$ basis, Alice and Bob start with
\begin{equation}
    \ket{\psi_{A/B}}=\frac{1}{\sqrt{2}}(\ket{0}\ket{+_\alpha} + \ket{1}\ket{-_\alpha}),
\end{equation}
where $\ket{\pm_\alpha} = e^{-\abs{\alpha}^2} \sum_{n=0}^\infty \frac{\alpha^n}{\sqrt{n!}}(a^\dagger/\sqrt{2} \pm b^\dagger/\sqrt{2})^n\ket{0}$. 
\\
As the server sends out a Bell state of the form $\ket{\Phi^+}$, we again perform a BSM with photon-number-resolving detectors, resulting in a perfect Bell state between the register $A$ ($B$) and the node $S_A$ ($S_B$). After this, entanglement can be swapped throughout the repeater chain, creating entanglement between $A$ and $B$. Similarly to the single click case, we can implement the decoy state method in the test basis, which in this case is the X basis.
\end{document}